\documentclass[fleqn,usenatbib]{mnras}

\usepackage{newtxtext,newtxmath}
\usepackage[T1]{fontenc}

\DeclareRobustCommand{\VAN}[3]{#2}
\let\VANthebibliography\thebibliography
\def\thebibliography{\DeclareRobustCommand{\VAN}[3]{##3}\VANthebibliography}

\usepackage{graphicx}	% Including figure files
\usepackage{amsmath}	% Advanced maths commands
\usepackage{anyfontsize}
\usepackage{multirow}
\usepackage{cleveref}
\usepackage[starfontserif]{starfont}

\newcommand{\wotan}{{\tt w\={o}tan}}
\newcommand{\monocbp}{{\tt mono-cbp}}
\newcommand{\tess}{\textit{TESS}}
\newcommand{\kepler}{\textit{Kepler}}
\DeclareSymbolFont{starfontsym}{OT1}{sts}{m}{n}
\DeclareMathSymbol{\mathTerra}{\mathord}{starfontsym}{76}

\title[Finding Circumbinary Planets]{Finding Circumbinary Planets: A Semi-Automated Transit Search of \tess\ Eclipsing Binaries}

\author[B. D. R. Davies et al.]{
Benjamin D. R. Davies,$^{1,2}$\thanks{E-mail: ben.d.r.davies@warwick.ac.uk}
David J. A. Brown,$^{1,2}$
Samuel Gill,$^{1,2}$
Jenni R. French$^{3}$
\\
$^{1}$Department of Physics, University of Warwick, Gibbet Hill Road, Coventry, CV4 7AL, UK\\
$^{2}$Centre for Exoplanets and Habitability, University of Warwick, Gibbet Hill Road, Coventry, CV4 7AL, UK\\
$^{3}$School of Physics and Astronomy, University of Birmingham, Edgbaston, Birmingham, B15 2TT, UK\\
}

\date{Accepted 2026 April 09. Received 2026 April 09; in original form 2025 October 02}

\pubyear{\the\year{}}

\begin{document}
\label{firstpage}
\pagerange{\pageref{firstpage}--\pageref{lastpage}}
\maketitle

% Abstract of the paper
\begin{abstract}
The discovery of circumbinary planets (CBPs) has advanced our understanding of planet formation and dynamical evolution in complex environments. However, the population of such planets remains small, leading their underlying physical properties to be loosely constrained. In this work, we have developed a semi-automated framework to identify planetary transit events in light curves of eclipsing binaries observed by the \textit{Transiting Exoplanet Survey Satellite} (\tess). Our search method, \monocbp, removes stellar eclipses and applies a custom detrending procedure, searching for individual transit events and applying automated vetting procedures to filter false positive signals. We searched a sample of binaries from the \tess\ Eclipsing Binary Catalogue, yielding one candidate transit event. \monocbp\ was also tested on the known population of transiting CBPs, using the \kepler\ long-cadence photometry for the \kepler\ transiting CBPs and the \tess\ Full Frame Image photometry for the \tess\ CBPs. Excluding transits that are shallower than the intrinsic noise of the \kepler/\tess\ data, \monocbp\ achieved a recovery rate of $\geq50$ per cent for each planet, reaching $>75$ per cent for 9 of the 14 planets. To test the limits of our framework, we injected simulated transit profiles with varying depth and duration into our sample of \tess\ light curves, finding that our recovery rate is a strong function of transit duration and the metrics used to filter false positive signals. This framework may be applied to large samples of \tess\ eclipsing binaries with little computational burden and to photometry from future space-based photometric surveys.
\end{abstract}

% Select between one and six entries from the list of approved keywords.
% Don't make up new ones.

%%%%%%%%%%%%%%%%%%%%%%%%%%%%%%%%%%%%%%%%%%%%%%%%%%

%%%%%%%%%%%%%%%%% BODY OF PAPER %%%%%%%%%%%%%%%%%%

\section{Introduction}
\label{sec: intro}

While many thousands of exoplanets have been discovered in recent decades, circumbinary planets (CBPs; planets that orbit the centre of mass of a stellar binary) make up only a small proportion of them. Before the first exoplanets were discovered, \citet{borucki_photometric_1984} suggested targeting eclipsing binaries (EBs) for transiting exoplanet surveys; EBs by definition have inclinations close to $90^{\circ}$ and, assuming that CBPs form in circumbinary discs which are coplanar with the orbital plane of the stellar binary, CBPs have an increased transit probability compared to planets orbiting single stars. The detection of these planets has only been possible, however, since the advent of space-based photometric surveys, in particular the \kepler~mission~\citep{Borucki2010}, which made possible the discovery of 12 transiting CBPs~\citep{doyle_kepler-16_2011, orosz_kepler-47_2012, orosz_neptune-sized_2012, orosz_discovery_2019, welsh_transiting_2012, welsh_kepler_2015, kostov_gas_2013, kostov_kepler-413b_2014, kostov_kepler-1647b_2016, schwamb_planet_2013}. The \kepler\ discoveries led to refinements of planet formation and migration theories in binary star systems~\citep[e.g.,][]{paardekooper_how_2012,penzlin_parking_2021,coleman_global_2023} and indicated that giant planets are likely to be as common, if not more so, than giant planets in single star systems (${\sim10}$ per cent for ${R_{\rm{p}}>6\rm{R}_{\mathTerra}}$~\citep{armstrong_abundance_2014}, or ${\sim9}$ per cent for ${M_{\rm{p}}>0.15\rm{M_{J}}}$~\citep{martin_planets_2014}).

Due to the gravitational potential of the inner binary, the orbits of CBPs are only stable beyond a critical semi-major axis $a_{p, \text{crit}}$~\citep[$\approx2-4a_b$;][]{Dvorak1984,Holman1999,Quarles2018}, where $a_b$ is the semi-major axis of the stellar binary. This stability constraint biases CBPs towards longer orbital periods compared to the current population of transiting planets around single stars. Therefore, to maximise efficiency for transiting CBP detection, a long, continuous time baseline of observations is preferred. A given CBP also produces shallower transits than the same planet transiting a single star, owing to contamination from the star not being transited. These challenges were overcome by the \kepler\ mission, with its 4 year near-continuous observational baseline and high-precision photometry both well-suited to discovering transiting CBPs.

Searching for transiting CBPs using NASA's \textit{Transiting Exoplanet Survey Satellite} \citep[\tess;][]{ricker_transiting_2015} is more challenging. \tess\ observes the sky in $\sim27$-day sectors, each covering a $24^\circ\,\times\,96^\circ$ region, before moving to an adjacent sector to tile the ecliptic hemispheres. With the exception of the Continuous Viewing Zones (CVZs), which receive $\sim1$ year of continuous photometry, this observing strategy, compared to \kepler, does not provide long durations of continuous photometry. \tess\ also does not provide the same level of photometric precision as \kepler. However, it does monitor a much larger number of stars, which are in general much brighter than the \kepler\ targets, hence any CBPs that are discovered by \tess\ are well suited for follow-up observations, with the potential for more detailed characterisation of individual systems than the \kepler\ sample \citep[e.g.,][]{Standing2023,Sairam2024}. To date, \tess\ has detected two transiting CBPs \citep[TOI-1338\,b and TIC 172900988\,b;][]{kostov_toi-1338_2020,kostov_tic_2021}.

The population of transiting CBPs, while small in number, have uncovered several interesting trends which have informed theories of the formation and evolution of binary stars and CBPs. Firstly, the majority of the known transiting CBP systems have a planet orbiting close to the inner stability limit of their stellar binary hosts. This has been inferred to have a physical origin rather than being a result of the observational bias of the transit method towards shorter period planets~\citep{martin_planets_2014,Li2016}, although the significance of this feature depends on how one defines the proximity of a CBP to the stability limit~\citep{Quarles2018}. This pile-up has been explained by inward migration of CBPs in the circumbinary protoplanetary disc after formation at larger radial distances, with the planet ceasing migration near the inner cavity of the disc, which coincides with the inner orbital stability limit for binaries with small eccentricities~\citep{Pierens2013,Kley2014,penzlin_parking_2021,Coleman2022}.

Secondly, there have been no CBPs detected orbiting short-period binaries ($<7$\,d), despite the overabundance of EBs with these shorter periods relative to EBs with periods $>7$\,d and the tighter inner stability limit allowing for stable orbits at shorter periods~\citep{armstrong_abundance_2014,martin_planets_2014,Li2016}. One channel for the formation of such tight binaries is thought to arise from the presence of tertiary companions. The theory proposes that these binaries initially formed with much larger orbital separations and subsequently underwent Kozai-Lidov oscillations induced by a tertiary stellar companion, with tidal forces dissipating the orbital angular momentum of the inner binary~\citep{Mazeh1979,Fabrycky2007}. This is supported observationally, with 96 per cent of stellar binaries with orbital periods $<3$\,d having a tertiary stellar companion, the same being true for only 34 per cent of binaries with periods $>12$\,d~\citep{Tokovinin2006}. The dynamics of such triple stellar systems restrict the parameter space for stable orbits of CBPs, particularly those that are more easily detectable~\citep[i.e., giant planets that are coplanar with the inner binary;][]{Martin2015,Munoz2015,Hamers2016}. There are also other proposed explanations for the lack of observed CBPs around short-period binaries, such as coupled stellar-tidal evolution~\citep{Fleming2018} as well as the reduction in transit probability due to tidal shrinkage of the stellar binary orbit~\citep{Mogan2025}.

A particular challenge of detecting transiting CBPs is their aperiodic transits, due to strong transit timing variations (TTVs). The largest contribution to this phenomenon is the time-dependent geometry of the stellar binary; the binary orbital motion causes differences in the position of the star and planet at the time of conjunction between successive orbits. CBPs also experience significant dynamical TTVs on multiple timescales. On short timescales, the non-axisymmetric quadrupole potential of the stellar binary induces oscillations in the planet's osculating period and eccentricity on a timescale of $\sim P_{\rm{bin}}$/2, producing TTVs on the order of hours. Over longer timescales ($\sim10-100$ years for the known transiting CBPs), apsidal and nodal precession generate substantially larger TTVs with amplitudes of several days. In some cases, nodal precession can cause transits to stop occurring for years~\citep{Schneider1994}, as is the case for Kepler-413\,b~\citep{kostov_kepler-413b_2014}. Due to these phenomena, many common transit search algorithms that assume strictly periodic planetary orbits, such as Box Least Squares~\citep[BLS;][]{kovacs_box-fitting_2002} or Transit Least Squares~\citep[TLS;][]{hippke_optimized_2019}, are ineffective for detecting transiting CBPs.

Transiting CBPs also exhibit strong transit duration variations (TDVs). Unlike planets on circumstellar orbits, a CBP's transit duration depends on the relative tangential velocities of the planet and the transited star, which varies with the binary's orbital phase. When the planet and star move in the same direction (near secondary eclipse for a prograde orbit), the planet takes longer to cross the stellar disc, producing an extended transit. Conversely, when they move in opposite directions (near primary eclipse), the transit is shorter in duration. For example, Kepler-47\,b exhibits transit durations ranging from $\sim3$-10 hours~\citep{orosz_kepler-47_2012,orosz_discovery_2019}. These large variations in transit duration present an additional challenge for automated detection algorithms for transiting planets, which typically assume a consistent transit shape and duration.

The majority of the \kepler\ and \tess\ transiting CBPs were discovered by visually inspecting light curves of EBs. However, several automated search methods have been developed. One approach has been to directly correct for the TTVs, which is achieved by using Keplerian models to approximate binary and CBP orbits~\citep[e.g.,][]{ofir_algorithm_2008,windemuth_automated_2019}, or by using N-body integrations~\citep{martin_searching_2021}. While these methods have been successful in detecting the known \kepler\ transiting CBPs and placing constraints on other planets in these systems, they can be computationally intensive, particularly for a blind search of a large sample of EBs. They also assume knowledge of the binary parameters\footnote{\citet{martin_searching_2021} note that their search method can treat the binary properties as free parameters, but the authors do not recommended this for a blind search due to long computational times.}, particularly stellar masses, which typically require spectroscopic follow-up\footnote{However, we note that it is possible to obtain estimates for EB stellar masses without radial velocities~\citep[e.g.,][]{Windemuth_modelling_2019}.}.

Another approach has been to perform a quasi-periodic transit search without predicting the transit times of potential CBPs, and therefore without the need for the host stellar masses. For example, \citet{kostov_gas_2013} developed a method to search for individual transits of CBPs by duplicating sections of \kepler\ EB light curves, performing a BLS search at the duration of the light curve segment and comparing transit and ``anti-transit'' events. \citet{armstrong_abundance_2014} identified individual transit events in \kepler\ EB light curves and performed a quasi-periodic transit search within the maximum TTV window defined in \citet{armstrong_placing_2013}. \citet{Klagyivik2017} fit a string of box-shaped transits to \textit{CoRoT} EB light curves, sampling a grid of planet periods, transit depth and duration, and allowing the transit times to vary. These search methods may lead to a greater number of false positive detections, but provide a bit more flexibility and are less computationally intensive.

In this work, we present a similarly flexible method based on those used by searches for long-period planets and eclipsing binaries~\citep[e.g.,][]{osborn_single_2016,gill_ngts_2020, gill_long-period_2020,gill_ngts-11_2020,grieves_old_2022,hawthorn_tess_2024}, namely searching for single-transit events (monotransits). This has the advantage of being effective at flagging candidates for follow-up observations in photometry with many gaps. It also has the ability to identify ``one-two punch'' events (when a CBP transits both components of the stellar binary in the same conjunction \citep{kostov_multiple_2020}), since there are no prior assumptions on periodicity or whether the planet transits the primary or secondary star. We apply our method to search a collection of \tess\ EB light curves for new transiting CBP candidates. 

Section~\ref{sec: framework} gives an overview of our framework to identify and vet individual transit events. Section~\ref{sec: data} describes the sample that we used for our transit search, taken from the \tess\ Eclipsing Binary Catalogue~\citep[][hereafter TEBC]{prvsa_tess_2022}. Section~\ref{sec: results} presents the results of our search, while Section\,\ref{sec: known CBPs} describes the performance of the framework on the known transiting CBPs. We then performed injection-recovery tests to assess the detection limits of our search in Section~\ref{sec: detection}. We discuss our findings and future directions in Section~\ref{sec: discussion} and present our conclusions in Section~\ref{sec: conclusions}.

\section{Detection Framework}
\label{sec: framework}

In this Section, we describe our semi-automated framework for identifying transits of CBPs, starting from raw photometry of EBs with known ephemerides and eclipse parameters to a candidate list of additional transit events. Our detection framework, which we dub \monocbp\footnote{https://github.com/bdrdavies/mono-cbp}, consists of four main steps: masking of stellar eclipses (Section~\ref{subsec: eclipses}), removing astrophysical and systematic trends from the light curves (Section~\ref{subsec: detrending}), searching for threshold crossing events (TCEs, Section~\ref{subsec: search}), and candidate vetting (Section~\ref{subsec: vetting}).

\subsection{Masking stellar eclipses}
\label{subsec: eclipses}

Typically, CBP transits have depths which are orders of magnitude smaller than the eclipses of a stellar binary, with the latter dominating the morphology of an EB light curve. Therefore, to make CBP transits easier to identify, the stellar eclipses must be removed.

There are two main ways to do this. Firstly, one could model the stellar eclipses and subtract this model from the light curve. This would ensure that residual in-eclipse data are passed to the transit search, enabling more efficient detection of syzygys (blending between transits and stellar eclipses). However, this becomes computationally cumbersome for large datasets, and any imperfection in the model can create transit-like features in the photometry after the model is subtracted, causing an increase in the rate of false positives. Additionally, the depth of stellar eclipses may change between sectors due to a variety of instrumental effects, primarily due to changes in the level of dilution from nearby sources and scattered light as the spacecraft changes orientation, requiring a separate light curve model for each sector and target.

Alternatively, one could identify in-eclipse data and mask it from subsequent stages in the transit search. This has the advantage of being much more computationally tractable if one already possesses the binary ephemeris and eclipse widths/positions, but significantly reduces sensitivity to syzygys. Due to the computational challenges of the former method, we instead opted to mask stellar eclipses from the light curves.

To mask the eclipses of a given EB, the ephemeris of the binary is used to phase-fold the light curve and removed all epochs within $\pm w_{\rm{p,s}}/2P_{\rm bin}$ of the eclipse mid-point, where $w_{\rm{p,s}}$ are the widths of the primary and secondary eclipses. The in-eclipse data is then masked for all subsequent steps of the framework. For \monocbp, these parameters (eclipse widths, period, reference epoch, and phase of secondary eclipse) are required as input.

\subsection{Detrending}
\label{subsec: detrending}

Removing trends from time-series photometry (or ``detrending'') is a vital step when searching for transiting exoplanets. Stellar variability (such as starspot modulation, flares, and pulsations), as well as instrumental systematics (e.g., pointing jitter, scattered light, and momentum dumps), can obscure or mimic real transit signals. Therefore, by removing these trends from light curves, one increases the detection sensitivity and reduces the number of false positive signals of their search.

For an automated, blind search of many light curves, it becomes infeasible to manually devise a detrending approach for each individual light curve, necessitating a general detrending approach that can be applied to light curves with a wide range of stellar and instrumental variability. In addition, one does not know \textit{a priori} what the properties of the transits they are searching for are, especially in the case of CBPs where there are significant TDVs. Additionally, EB light curves tend to exhibit more stellar variability than single stars due to a plethora of physical effects such as ellipsoidal modulation, reflection, Doppler beaming, and starspot modulation contributions from both host stars~\citep{Strassmeier2009,Shporer2017,Lurie2017}.

Due to its flexibility and range of available detrending algorithms, the \wotan\ software package~\citep{hippke_wotan_2019} was chosen to detrend the light curves before passing them to the transit search algorithm. Our general detrending approach, similar to and inspired by~\citet{martin_searching_2021}, is comprised of two steps: iterative cosine detrending and biweight detrending.

\subsubsection{Iterative cosine detrending}
\label{subsubsec: cosine}

The first stage of the detrending accounts for sinusoidal, quasi-periodic variability inherent to many EB light curves. Since these effects are not seen in all EB light curves, we first test for significant periodicity by calculating the Lomb-Scargle periodogram of the light curve (after masking eclipses). If there is a peak in the periodogram with a False Alarm Probability (FAP) of <1 per cent within the frequency range [1/27,2]\,d$^{-1}$, we fit the data using the {\tt cosine} method of \wotan~\citep[see section 2.2 of][]{Mazeh2010} with a window length of 12\,d (defined based on testing using the known transiting CBPs) and recalculate the periodogram. If there are any peaks within the above frequency range that remain after this trend is removed that have a FAP below our 1 per cent threshold, the raw data is fit with the {\tt cosine} method with iteratively decreasing window length (increments of 0.1\,d, with a minimum of 1\,d), recalculating the Lomb-Scargle periodogram for each fit. This process continues until there are no longer any peaks in the Lomb-Scargle periodogram with a FAP below 1 per cent.

\subsubsection{Biweight detrending}
\label{subsubsec: biweight}

To remove any remaining trends, we applied a second detrending step which uses the \wotan\ implementation of Tukey's biweight time-windowed slider~\citep{Mosteller1977}. This was shown by~\citet{hippke_wotan_2019} to be the most effective method of detrending light curves for transit detection, in terms of preserving the transit signal.

Our approach to this second detrending step is similar to~\citet[][see their section 2.1.3]{Devora-Pajares2024}. We define a grid of window lengths to pass to the biweight detrending function, which we chose to be [1,3]\,d with a step size of 0.1\,d. Then, for each window length, the \wotan\ {\tt flatten} function is applied to the cosine-detrended light curve, yielding a total of 21 light curves which have been detrended using different biweight window lengths. If a given event is identified by the search algorithm (Section~\ref{subsec: search}) in a specified proportion of these biweight-detrended light curves, the event is flagged as ``detrending-independent''. Events that are identified in fewer than this proportion of light curves are ``detrending-dependent'', and hence more likely to be an artefact of poor detrending. For our purposes, we found that a threshold of 18/21 light curves for this metric was a suitable value to exclude detrending artefacts while still retaining known transit signals. Therefore, if an event is detected in 18 or more of the 21 light curves, it is classed as detrending-independent, and if an event is detected in 17 or fewer of the 21 light curves, it is classed as detrending-dependent.

We note that while 1\,d detrending windows for both the cosine and biweight detrending risk reducing the signal-to-noise ratio (SNR) of long-duration CBP transits, we found that they were necessary to adequately remove red noise from more variable sources, where such transits would otherwise be difficult to detect.

\subsection{Transit search}
\label{subsec: search}

The method we use to identify single transit events is similar to that used in~\citet{hawthorn_tess_2024}. Firstly, at each cadence, the median absolute deviation (MAD) of a 4\,d window of each detrended light curve is calculated (after the stellar eclipses are masked). Then, the algorithm flags a TCE if 3 consecutive cadences lie below a threshold determined by the MAD (typically 3\,$\times\,$MAD). This method was designed for the 30 minute cadence Full Frame Image (FFI) photometry, hence all data with a faster cadence is binned to 30 minutes before passing it to the search algorithm.

This component of the framework also estimates the duration, $t_{\text{dur,mono}}$, and depth, $\delta_{\text{mono}}$, of the event. $t_{\text{dur,mono}}$ is defined as the difference between the times before and after the event when the normalised flux rises above 1. $\delta_{\text{mono}}$ is defined as the median of the normalised flux within $t_{\text{dur,mono}}$.

Our search method is effective at flagging events in sparse photometry, and is capable of identifying ``one-two punch'' events. However, it does require transits with large enough SNR to be individually discernable and, due to the lack of physical constraints on detections, there is a risk of many false positive detections~\citep[see e.g.][]{hawthorn_tess_2024}.

\subsection{Vetting}
\label{subsec: vetting}

Our detection technique is agnostic to both the periodicity of events and their morphology. This has the advantage of reliably identifying high SNR monotransits, but the majority of features in a light curve that satisfy these search criteria and survive the detrending process are systematic trends and artefacts left over from poor detrending. In an attempt to filter out these false positive signals, we developed a series of vetting steps.

\begin{figure*}
 \includegraphics[]{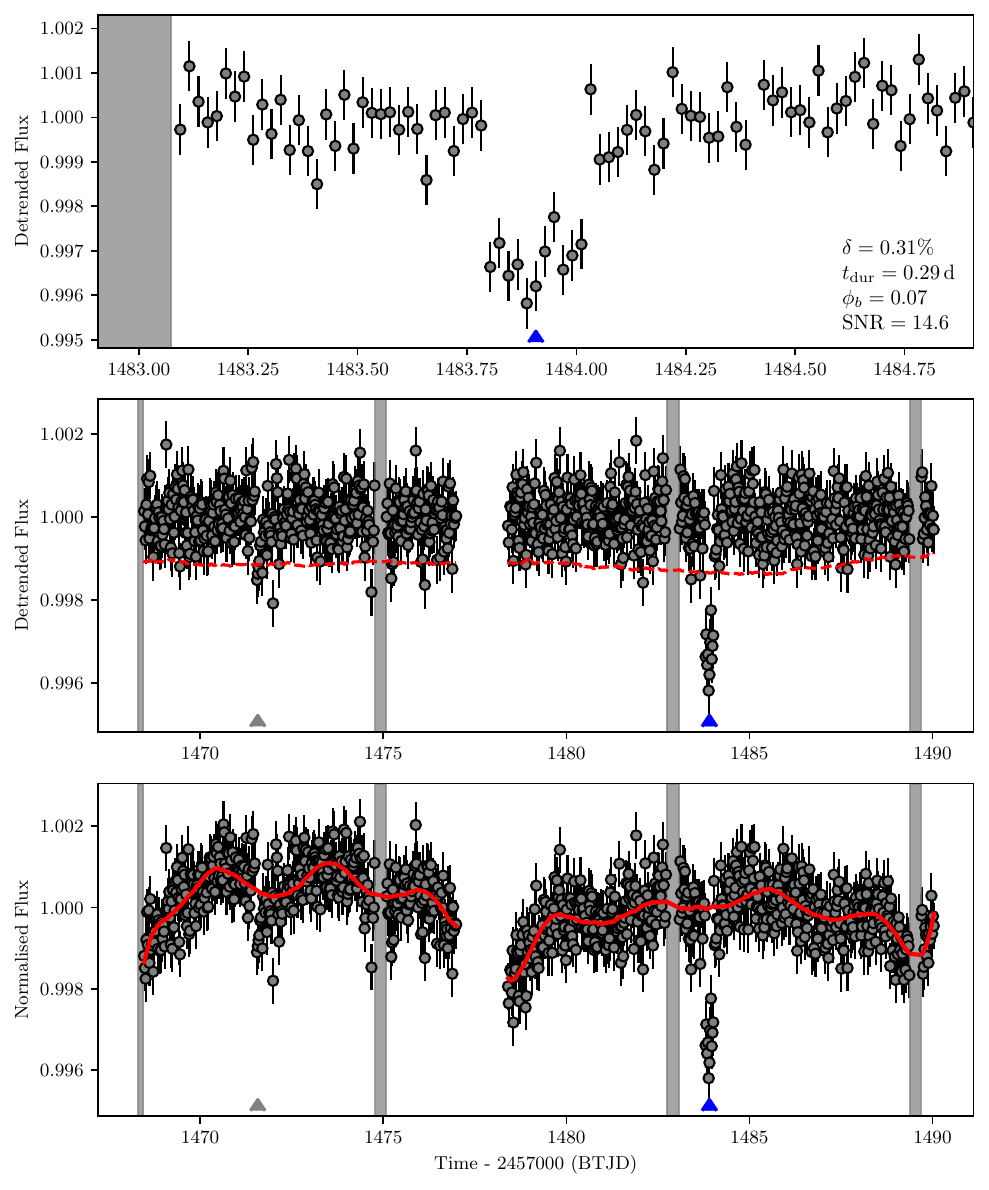}
 \caption{The detection of a transit of TOI-1338\,b by \monocbp. \textbf{Top:} A zoom-in of the identified event. The grey points depict the detrended flux. The legend shows the estimated parameters, including the depth ($\delta$), duration ($t_{\text{dur}}$), binary phase at the time of the event ($\phi_b$), and the event SNR (see Eq.~\ref{eq: SNR}). \textbf{Middle:} The full detrended light curve. \textbf{Bottom:} The full raw light curve. The solid red curve shows the detrending model fit to the data, and the red dashed line shows our detection threshold. \textbf{All panels:} The carets point out the location of all TCEs identified in the light curve, with the blue caret pointing out the transit event. The grey regions indicate epochs corresponding to the primary and secondary eclipses of TOI-1338AB, which were masked prior to the transit search.}
 \label{fig: example vetting plot}
\end{figure*}

Once the transit search has been performed, the time of each event, as well as the \tess\ Input Catalogue (TIC) ID and sector, is recorded and output vetting plots are created which may be used for visual inspection. An example of these vetting plots for one of the transits of TOI-1338\,b is shown in Fig.~\ref{fig: example vetting plot}. These plots are helpful for checking the performance of the detrending algorithm, as well as for providing a quick view of the event and its properties.

\monocbp\ also applies some automated vetting checks using a series of flags and thresholds. Firstly, the Skye excess metric \citep[see e.g.][]{thompson_planetary_2018,Fernandes2022} is used to determine if an event is correlated in time with events occurring on other targets, indicative of an instrumental systematic. For each \tess\ sector, we group events that occur within 0.1\,d of each other. If the number of TCEs in a given group of events is an outlier across the sector (at $>3\sigma$), these TCEs are flagged as likely systematics. We note that it is possible that a real transit signal may coincide with such a systematic and be flagged as a false positive using this methodology. However, it is an effective method of reducing the number of false positive signals identified by the transit search (see Section~\ref{subsec: FPR}).

\begin{table}
 \caption{Models that are fit to each event for the model comparison of \monocbp. Each fitted parameter is explained in detail in Section~\ref{subsec: vetting}.}
 \label{tab: models}
 \begin{tabular}{lll}
  \hline
  Model & Fitted Parameter & Prior Distribution\\
  \hline
  Transit & Median Flux & $\mathscr{N}$(1, 0.01)\\
   & $t_{0}$ & $\mathscr{N}$($t_{0,\text{mono}}$, 0.05)\,d\\
   & $q_{1}$ & $\mathscr{U}$(0, 1)\\
   & $q_{2}$ & $\mathscr{U}$(0, 1)\\
   & $\delta$ & $\mathscr{N}$($\delta_{\text{mono}}$, $\delta_{\text{mono}}/2$)\\
   & $b$ & $\mathscr{U}$(0, 1)\\
   & $t_{\text{dur}}$ & $\mathscr{N}$($t_{\text{dur, mono}}$, 0.1)\,d\\
  Sinusoid & Median Flux & $\mathscr{N}$(1, 0.01)\\
   & $A$ & $\mathscr{N}$($\text{max}(F)-\text{min}(F)$, 0.1)\\
   & $\phi$ & $\mathscr{U}$(0, $2\pi$)\\
   & $\omega$ & $\mathscr{U}(1/t_{\text{base}}, 2/t_{\text{base}})$\\
  Linear & $m$ & $\mathscr{N}$($m_{\text{guess}}$, 0.1)\\
   & $c$ & $\mathscr{N}$($c_{\text{guess}}$, 0.1)\\
  Step & $a_1$ & $\mathscr{N}$($a_{1,\text{guess}}$, 0.1)\\
   & $b_1$ & $\mathscr{N}$($b_{1,\text{guess}}$, 0.1)\\
   & $c_1$ & $\mathscr{N}$($c_{1,\text{guess}}$, 0.1)\\
   & $a_2$ & $\mathscr{N}$($a_{2,\text{guess}}$, 0.1)\\
   & $b_2$ & $\mathscr{N}$($b_{2,\text{guess}}$, 0.1)\\
   & $c_2$ & $\mathscr{N}$($c_{2,\text{guess}}$, 0.1)\\
  \hline
 \end{tabular}
\end{table}

Secondly, a SNR cut is applied to the identified events, where we define the SNR of a single-transit event as

\begin{equation}
    \text{SNR}=\frac{\delta}{\bar{\sigma}}\sqrt{N_{\rm data}},
    \label{eq: SNR}
\end{equation}

\noindent where $\delta$ is the depth of the event, and $N_{\rm data}$ is the number of data points within the event, defined by $t_{\text{dur}}/\delta t$, where $t_{\rm dur}$ is the duration of the event and $\delta t$ is the cadence of the data, both in units of days. We define  $\bar{\sigma}$ as a linear combination of the median flux uncertainty within the duration of the event and the local flux scatter, the latter given by the standard deviation of out-of-event flux in a window centred on the event. The size of this window is defined as $3t_{\text{dur}}$ or 1\,d, whichever is longer. In order to be as inclusive as possible, we decided to apply a threshold of $\text{SNR}>5$ for building our candidate list. This threshold was chosen since it corresponds to the lowest mean SNR of the known transiting CBPs (Kepler-47 b, see Fig.~\ref{fig: detection efficiency known CBPs}), and to minimise the risk of removing real transit signals from our search. Since each light curve is detrended using using multiple biweight window lengths (see Section~\ref{subsubsec: biweight}), \monocbp\ may estimate slightly different depths and durations (and hence different SNRs) for the same event across these detrended versions. Therefore, we chose the light curve where a given event has the highest SNR to derive the final recorded properties of the event.

In addition, we apply a limit on the event duration of $<1$\,d, since we find that systematic trends and detrending artefacts are more likely than CBP transits to have such a long duration. We also consider only those events that are detrending-independent (see Section~\ref{subsubsec: biweight}).

\begin{figure*}
 \includegraphics[]{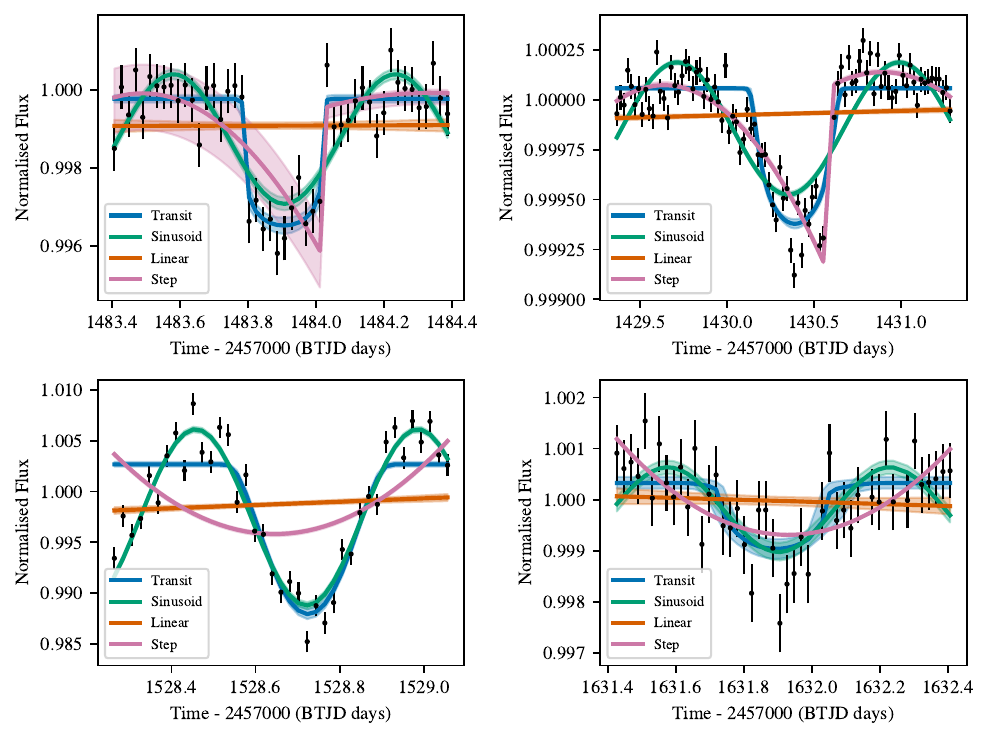}
 \caption{Examples of events identified by \monocbp\ and vetted by the model comparison (Section~\ref{subsec: vetting}). \textbf{Top left:} Sector 6 transit of TOI-1338 b (same as Fig.~\ref{fig: example vetting plot}), classified as a transit. \textbf{Top right:} An instrumental systematic occurring in the Sector 4 light curve of TIC 55497281, classified as a step. \textbf{Bottom left:} Poor removal of stellar variability in the Sector 8 light curve of TIC 5092088, classified as a sinusoid. \textbf{Bottom right:} An unclassified event occurring in Sector 12 light curve of TIC 260502102, where there was no preferred model. In all panels, the shaded regions depict the 16th and 84th percentiles of the posterior distributions for each model.}
 \label{fig: model comparison example}
\end{figure*}

For our final vetting stage, we fit a series of models to each event that passes the previous vetting stages. The specifics of each model, including fitted parameters and prior probability distributions, are outlined in Table~\ref{tab: models}.

To fit the transit model, we use the mid-event time $t_{0,\text{mono}}$, event duration $~t_{\text{dur,mono}}$, and event depth $\delta_{\text{mono}}$ estimated in the transit search as priors, as well as wide uniform priors on the impact parameter $b$ and the quadratic limb-darkening parameters $q_1$ and $q_2$~\citep{Kipping2013}. The model used to evaluate the light curve of each event is described in~\citet{exoplanet:agol20}, implemented within {\tt exoplanet}~\citep{Foreman-Mackey2021}.

For the sinusoidal fit, we fit each event to a simple sinusoidal model with amplitude $A$, phase $\phi$, frequency $\omega$, and median flux $\bar{F}$, given by $A\sin{(2\pi\omega t_{\text{event}}+\phi)}+\bar{F}$. We sample a uniform distribution for $\omega$ from $1/t_{\text{base}}$ to $2/t_{\text{base}}$, where $t_{\text{base}}$ is the time baseline of the window used to calculate $\bar{\sigma}$.

The step fit tests whether there is a significant jump in flux. The difference in flux between successive cadences in the event window is calculated. If the maximum flux difference is a $3\sigma$ outlier, a 2nd order polynomial is fit to the data on either side of the jump as follows:

\begin{equation}
    F_{\text{step}}=\begin{cases}
    a_1t_{\text{event}}^2+b_1t_{\text{event}}+c_1, & \text{if $t_{\text{event}}\leq t_{\text{step}}$}.\\
    a_2t_{\text{event}}^2+b_2t_{\text{event}}+c_2, & \text{if $t_{\text{event}}>t_{\text{step}}$},
    \end{cases}
    \label{eq: step fit}
\end{equation}

\noindent where $t_{\text{step}}$ is the time at which the maximum flux difference occurs. If there is no statistically significant jump, a single 2nd order polynomial is fit across the whole event window. To set priors for the fit, we calculate ``guess'' parameters (denoted by $a_{1,2,\text{guess}}$, $b_{1,2,\text{guess}}$, and $c_{1,2,\text{guess}}$ in Table~\ref{tab: models}) using least-squares polynomial fits to either side of the flux jump (if present), or to the whole event window if no flux jump is present.

Finally, we fit a linear model to the event. Similar to the priors for the step fit, we calculate ``guess'' parameters (with gradient prior $m_{\text{guess}}$ and intercept prior $c_{\text{guess}}$) using a least-squares linear fit to the event window.

For each model, we sample the posterior distribution using an Markov chain Monte Carlo (MCMC) approach, performing a total of 2000 draws across four Markov chains, with the first 1000 draws discarded as burn-in. This setup was chosen to balance computation time with adequate exploration of the parameter space for each model.

We then compare two criteria to assess the quality of each fit to the data in the event window: the Akaike Information Criterion~\citep[AIC;][]{Akaike1998} and the root mean square error (RMSE) of the residuals. The AIC and RMSE values for each model are then compared, classifying the events into different tiers. If there is a model $m$ such that $\text{AIC}_{m}-\text{AIC}_{n}\leq-2~\forall~n\neq m$, then this event is classified as the corresponding model $m$. For example, in Fig.~\ref{fig: model comparison example}, for the top left panel, the transit model fit satisfies this criterion, and hence is classified as a transit. If the RMSE value of the residuals of this model fit is $\leq1.2$, then the event is still classified according to the best AIC value, but picks up an ambiguity flag. If there is no preferred model based on the AIC comparison, the event is ``unclassified''.

This part of the framework also generates vetting plots that show the model fits to each event. Some examples of events with different classifications are shown in Fig.~\ref{fig: model comparison example}. The top-left panel shows the same transit event for TOI-1338\,b as in Fig.~\ref{fig: example vetting plot}, which, as mentioned above, is labelled as a transit. The top-right panel shows an example of an event that is classified as a step, the bottom-left shows an event classified as a sinusoid, and the bottom-right shows an event which was unclassified.

\section{Data and Sample Selection}
\label{sec: data}

We applied our methodology outlined in the previous section to the publicly available TEBC~\citep{prvsa_tess_2022}. This catalogue was created based on 2-minute cadence data processed by the \tess\ Science Processing Operations Center (SPOC) from \tess' Primary Mission (the first 2 years of observations) and contains 4584 EBs. For each entry in the catalogue there are several parameters that were calculated by~\citet{prvsa_tess_2022} which aid in the eclipse masking process (see Section~\ref{subsec: eclipses}) including the period, reference epoch, and eclipse parameters (widths, depths, and positions; see section 3.5 of~\citet{prvsa_tess_2022} for a description of how these parameters were obtained). The TEBC also provides a morphology coefficient for each EB. This is a continuous variable between 0 and 1 that describes how detached the binary is, with 0 corresponding to the most detached systems and 1 to purely ellipsoidal variables (see section 4 of \citet{matijevivc_kepler_2012} for a detailed description).

To constrain the sample of EBs to those which are of interest for our transit search, several cuts were applied. Firstly, all EBs that were flagged as ambiguous or as having insufficient time coverage were removed from the sample (see section 3.4 of~\citet{prvsa_tess_2022} for a description of these flags), removing 588 entries.

Secondly, we restricted our sample to detached EBs due to the complex light curve morphologies inherent to semi-detached and contact binaries, which make transits much more difficult to detect. Therefore, we chose to only include systems with morphology coefficient $<0.5$~\citep[this is the suggested threshold for detached EBs;][]{matijevivc_kepler_2012}. This reduced the sample to 1790 EBs.

We also limited our sample to binaries which have a period of $>7$\,d. The reasons for enforcing this limit is twofold. Firstly, light curves of short-period EBs possess more intrinsic red noise than longer-period EBs, since tidal locking induces fast rotation, subsequently inducing increased levels of activity (e.g., starspots and flares). These effects, as well as increased prominence of binary-specific light curve modulation (see Section~\ref{subsec: detrending}), make detrending the light curves of short-period EBs particularly challenging. Secondly, there are both theoretical predictions and observational evidence for the lack of coplanar CBPs orbiting short-period binaries (Section~\ref{sec: intro}). The high rate of tertiary companions for short-period binaries~\citep{Tokovinin2006} suggests that the majority of binaries with period $<7$\,d in the TEBC have a tertiary stellar companion, leading to a probable lack of giant CBPs in aligned orbits for these binaries. There may, however, exist small CBPs with stable, misaligned orbits in these systems which may sparsely transit their host stars~\citep{Martin2015,Munoz2015,Hamers2016}. Since our detection method requires high SNR individual transits, we would be unlikely to detect these smaller CBPs. Following this period cut, our sample is reduced to 591 EBs.

\begin{figure}
 \includegraphics[]{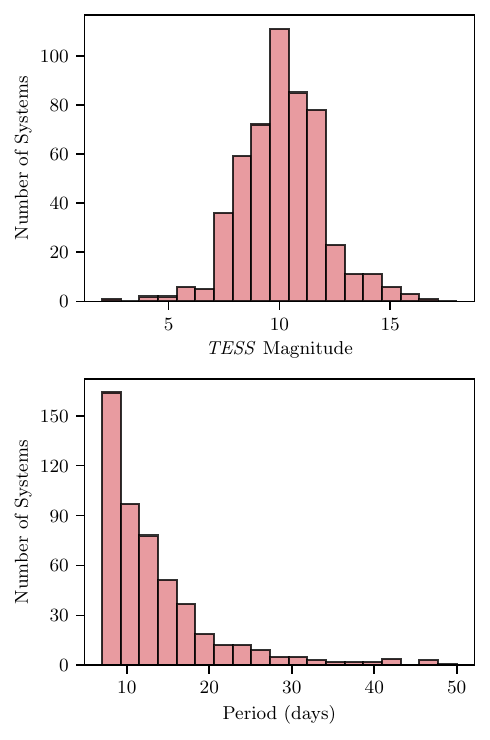}
 \caption{\textbf{Top}: \tess\ magnitude distribution of our sample of EBs. The median \tess\ magnitude is $\approx10.2$. \textbf{Bottom}: Period distribution of our sample of EBs. The median period in our sample is $\approx11.4$\,d. 6 EBs with periods $>50$\,d are not show in the period distribution for visualisation purposes.}
 \label{fig: EB sample}
\end{figure}

In addition to these cuts, we cross-matched the TIC IDs of the remaining 591 EBs with the TIC IDs of all published transiting planets in the NASA Exoplanet Archive\footnote{https://exoplanetarchive.ipac.caltech.edu/, accessed on 2025 March 31}~\citep{Christiansen2025}, returning 6 circumstellar planetary systems which are labelled as EBs in the TEBC: TOI-1339~\citep[HD 191939, TIC 269701147;][]{Badenas-Agusti2020}, TOI-677~\citep[TIC 280206394;][]{Jordan2020}, TOI-199~\citep[TIC 309792357;][]{hobson_toi-199_2023}, TOI-1736~\citep[TIC 408618999;][]{Martioli2023}, WASP-34~\citep[TIC 437242640;][]{Smalley2011}, and TOI-216~\citep[TIC 55652896;][]{kipping_resonant_2019}. These were removed from our sample.

As a final check, we visually inspected the phase-folded light curves of our remaining sample. We filtered out binaries that had no eclipse parameters, those with inaccurate ephemerides and eclipse parameters, those with no clear eclipses, and those which showed high-frequency ($>1\,\text{d}^{-1}$) pulsations. Our final sample consists of 512 EBs. The \tess\ magnitude and period distribution of this final sample is shown in Fig.~\ref{fig: EB sample}. The median \tess\ magnitude is 10.2 and the median period is 11.4\,d.

Initially, we took ephemerides from the TEBC, which were derived from early \textit{TESS} data. However, we find that many systems in our sample experience ephemeris decay, with the ephemerides from the TEBC no longer adequately fitting all of the photometry. This in turn leads to false positives from the transit search, as the eclipses are not correctly masked, leaving TCEs resulting from residual portions of the eclipse that are not removed during the masking step. For these systems, we employ a hybrid BLS and Phase Dispersion Minimisation (PDM) algorithm on the entirety of the \textit{TESS} data to recalculate the periods and ephemerides of systems exhibiting ephemeris decay (French et al. \textit{in prep}). We first search for periodicity with BLS, and then examine the residual lightcurve with PDM to find ephemeris decay and calculate updated ephemerides that account for this.

To perform our transit search, we chose to use the light curves produced by the TESS-SPOC pipeline~\citep{Caldwell2020}, which applies the SPOC data reduction pipeline~\citep{jenkins_tess_2016} to the \tess\ FFIs, producing the same format of data products as the 2-min cadence SPOC data. For our search, we use all of the available TESS-SPOC data at the time of writing (up to and including Sector 79, ending 2024 June 18). The 10-minute and 200-second cadence FFI light curves used in our search were binned to 30-minute cadence, necessitated by the design of our transit search (see Section~\ref{subsec: search}). We note that there are later sectors available for our targets provided by other data reduction pipelines, but we chose to use TESS-SPOC data for consistency across our sample. In addition, we applied the ``hard" quality flag filter from the {\tt lightkurve} software package~\citep{lightkurve_2018} to conservatively remove poor-quality data and data affected by stray light, thereby reducing the false positive rate.

The SPOC produces two photometric data products: Simple Aperture Photometry (SAP) and the Pre-search Data Conditioned Simple Aperture Photometry (PDCSAP). The SAP flux was preferred for our search owing to spurious transit-like features appearing in the PDCSAP flux that do not appear in the SAP flux, or from any other data reduction pipeline available on the Mikulski Archive for Space Telescopes (MAST). Additionally, for many light curves in our sample, the PDC process increases the scatter of the photometry, making the detection of CBP transit signatures more difficult. These effects may be due to the PDC process attempting to correct for light from contaminating sources in the aperture, occasionally resulting in overcorrection. Similarly, other searches for transiting CBPs in the \kepler\ data have preferred to use SAP over PDCSAP, and then perform their own detrending~\citep[see e.g.,][]{orosz_neptune-sized_2012,windemuth_automated_2019,martin_searching_2021}.

\section{Results of the TEBC search}
\label{sec: results}

We applied \monocbp\ to the sample of 512 EBs selected from the TEBC as outlined in Section~\ref{sec: data}. Here, we describe the outcome of this search.

\begin{figure}
 \includegraphics[width=\columnwidth]{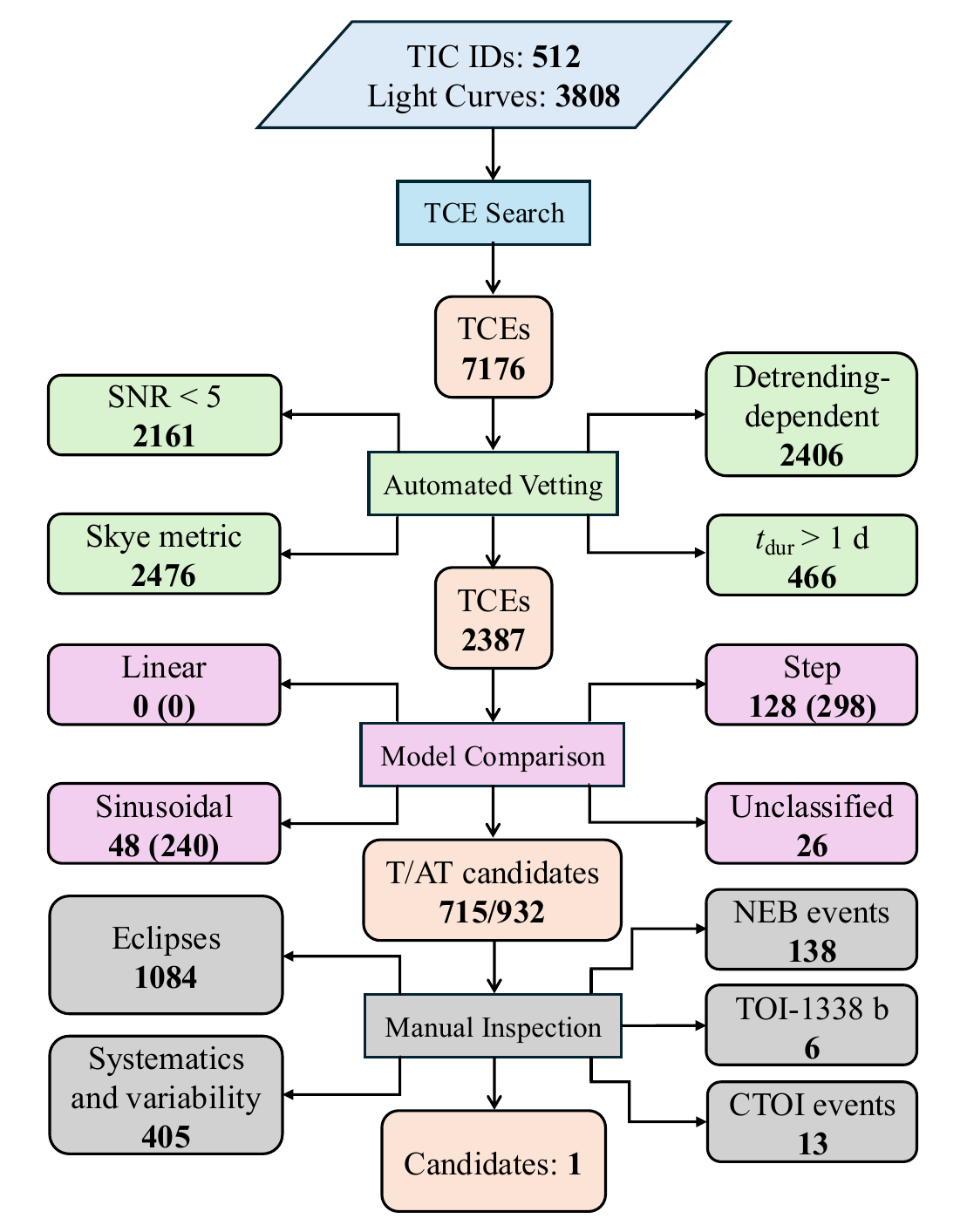}
 \caption{Overview of our search of the TEBC for candidate transit events. The number of TCEs that survive each step of the vetting process are denoted in the central column (peach boxes). The number of TCEs that are filtered by each step are shown in the boxes to the left and right of the central column. The automated vetting step is shown in green, model comparison in magenta, and the manual inspection component in grey. For the model comparison component, the values in parentheses denote the number of events flagged as ambiguous based on the RMSE criterion.}
 \label{fig: flowchart}
\end{figure}

\begin{figure*}
 \includegraphics[]{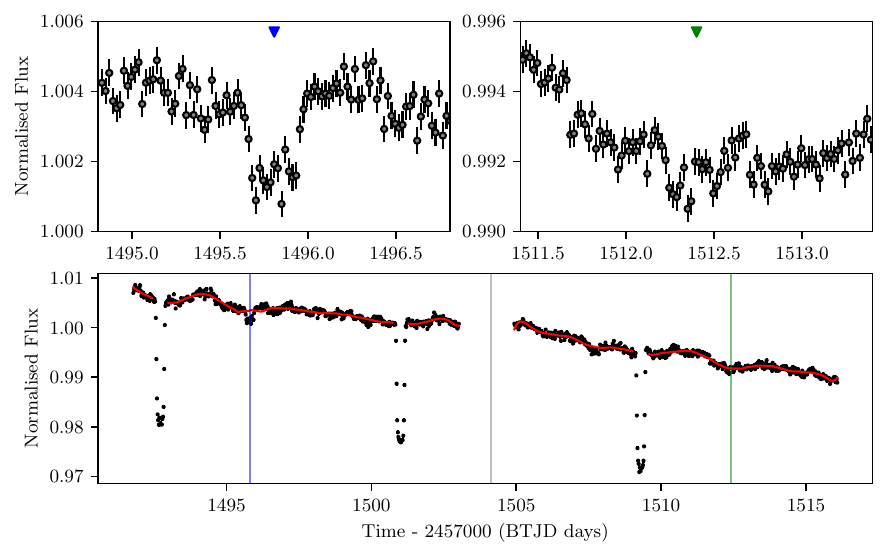}
 \caption{The candidate event identified in the Sector 7 TESS-SPOC light curve of TIC 319011894. \textbf{Bottom:} The Sector 7 TESS-SPOC light curve of TIC 319011894, with the \monocbp\ detrending depicted by a red curve. There are three eclipses with depth $\approx2.2$ per cent and period of 8.3\,d. The time of the candidate event ($\sim1495.8$ BTJD) is marked with a blue vertical line. If the candidate event was a secondary eclipse, there would be an event in the mid-sector data gap (grey vertical line) and an event at $1512$ BTJD (green vertical line). \textbf{Top left:} The candidate event. \textbf{Top right:} The flux when an observable secondary eclipse would occur, if the candidate event was a secondary eclipse. We point out that these measurements may have been affected by an instrumental systematic or scattered light, but we find no evidence of this from our vetting.}
 \label{fig: candidate event}
\end{figure*}

Figure~\ref{fig: flowchart} presents an overview of the number of TCEs that were identified in our search, and how these were processed through \monocbp. In summary, we applied \monocbp\ to 3808 TESS-SPOC FFI light curves (up to Sector 79) of the 512 EBs in our sample, identifying 7176 TCEs. After the default vetting metrics were applied (Skye metric flag, detrending-dependence flag, SNR and duration cuts; see Section~\ref{subsec: vetting} for details), we are left with a total of 2387 TCEs. After processing these remaining TCEs through the model comparison component of \monocbp, we identified a total of 1647 events that were classified as transits (T) or ambiguous transits (AT) for manual vetting. Note that the sum of the events that are filtered in the automated vetting stage (green boxes in Fig.~\ref{fig: flowchart}) are not equal to the total number of TCEs identified in the transit search since many spurious events are flagged with multiple metrics.

The majority of events were classified as transits or ambiguous transits ($\approx69$ per cent), implying that most of the events that pass the initial vetting checks have a transit-like morphology. Alternatively, it may imply that our transit model is more flexible than the other models that we use, although this is penalised by the AIC. Notably, there were no cases where the TCE was best described by the linear model.

\subsection{Manual vetting}
\label{subsec: candidates}

Despite the steps that we have implemented to filter false positives, they are not perfectly efficient and some false positive signals will remain. These may be astrophysical false positives, such as secondary eclipses that are not removed by the eclipse masking step, background variable sources, or residual stellar variability which was not removed by the detrending. Due to the large number of events that were classified as a transit or ambiguous transit by the model comparison, it can be inferred that there remain false-positive signals that survive the automated vetting steps and have transit-like morphologies. Therefore, to refine our candidate list, we manually vet each event that is classified as T or AT.

For all 1647 T/AT events, we visually inspected each vetting plot (e.g., Fig.~\ref{fig: example vetting plot}) to identify clear systematics and detrending artefacts, with the latter often being due to remaining signatures of stellar variability (e.g., pulsations and rotational modulation). Overall, these false positive T/AT events included 258 detrending artefacts and 135 systematics that had not been removed by the previous vetting steps.

There were also some events that were classified as T/AT by the model comparison which, upon further inspection, were due to contamination from a nearby, short-period eclipsing binary. These amounted to 138 T/AT events.

There are several binaries in our sample that do not have secondary eclipse parameters. For these targets, secondary eclipses are not processed by the eclipse masking procedure and so are flagged as TCEs. Additionally, since many of these have similar morphologies to planetary transits, most of these are classified as transits by the model comparison. If there are $\geq3$ events at the same binary phase for a given target, these are flagged as secondary eclipses. If this is not the case, we compare the phase of the event to the phase-folded light curve of the binary. If the phase of the event coincides with the phase of a visually-discernable eclipse, it is flagged as an secondary eclipse. Overall, 1084 T/AT events were identified as stellar eclipses.

Several of the 1647 T/AT TCEs had already been identified as Community Planet Candidates (CTOIs). For example, 6 events in the light curves of TIC 317015040 were flagged as transits by the model comparison (one in Sector 17, 57, 58, and 59, as well as two in Sector 18). One of the Sector 18 signals is associated with a planet candidate identified by the Xingming Obsevatory Exoplanet Search project (XESP\footnote{http://xjltp.china-vo.org/xesp.html}), with a period of $32.158\pm0.002$\,d. Upon inspecting the Target Pixel Files (TPFs) for TIC 317015040, the stellar eclipse signal is over an order of magnitude larger on TIC 317015048, originating 42.65 arcsec away from TIC 317015040, implying that the stellar eclipses are off-target. Additionally, even if the binary was on-target, the stellar eclipses have a period of 19.8\,d, leading to a period ratio $P_p/P_{\rm bin}\approx1.6$, well within the inner orbital stability limit. Therefore, we disregard this as a candidate CBP transit.

Additionally, 9 events in the light curves of TIC 261261490 were classified as either T or AT. In the TEBC, TIC 261261490 is listed as having a period of 113.0648\,d. These events correspond to a CTOI with a period $\approx3.5$\,d. Therefore, as with TIC 317015040, we rule out the possibility of these transits originating from a CBP.

Finally, 6 transits of TOI-1338\,b were identified and classified as T/AT. Since this is a known transiting CBP, we remove these from our final candidate list. In addition to our detected transits, TOI-1338\,b transits in Sectors 3, 30, 62, and 68. However, there is no TESS-SPOC light curve of TOI-1388 for Sector 3, and the other transits are all removed before the transit search since they were affected by stray light, and hence (at least partially) removed by the ``hard" data quality cut.

After performing these manual vetting checks, we refined our candidate list to 13 TCEs that required further investigation to confirm or rule out. For these events, we performed a separate series of manual checks, including comparing the SAP and PDCSAP photometry, checking the background for any correlated events in time (due to, e.g., an asteroid passing through the frame), checking all available light curves on MAST, checking for any correlated centroid shifts, and checking the light curves of each pixel in the SPOC aperture for correlated anti-transit events, caused by pointing jitter changing the shape of the point-spread function (PSF, see figures 3 and 4 of \citealt{kostov_toi-1338_2020}).

\subsection{A new CBP candidate}
\label{subsec: candidate}

Following these steps, we are left with a single candidate event. TIC 319011894 has a single sector of \tess\ data (Sector 7). It shows eclipses of depth $\approx2.2$ per cent with a period of 8.3\,d. There is an additional event at $1495.8$ BTJD with a depth of $\approx0.2$ per cent. At first, this was assumed to be a secondary eclipse. However, there does not appear to be another event at the corresponding phase of the binary orbit later in the sector (see Fig.~\ref{fig: candidate event}). With a period of $\sim8.3$\,d, TIC 319011894 may possess significant orbital eccentricity~\citep[e.g.,][]{VanEylen2016,IJspeert2024}, which could geometrically preclude observable secondary eclipses. Additionally, the morphology of the primary eclipses (i.e., they are flat-bottomed and shallow) indicates a low binary mass ratio, implying that any secondary eclipses would be shallow and potentially undetectable given the photometric precision.

\begin{table*}
 \caption{A selection of properties of the known transiting CBP systems. For systems with host binary mass ratio $q_{\rm bin}$ $\geq0.8$, the listed number of transits includes those across both primary and secondary stars, since these have similar depths. Note that the mean transit depth ($\bar\delta$) and the transit durations ($t_{\rm dur}$) were estimated from \monocbp's TCE search (Section~\ref{subsec: search}).}
 \label{tab: known CBPs}
 \begin{tabular}{lllllll}
  \hline
  CBP & $P_{\text{bin}}$ (d) & $q_{\text{bin}}$ & $\bar\delta$ (\%)$^\dagger$ & $t_{\text{dur}}$ (d) & $N_\text{tr}$ & Ref.\\
  \hline
  Kepler-16\,b & 41.08 & 0.29 & 1.80 & 0.31--0.43 & 7 & \citet{doyle_kepler-16_2011}\\
  Kepler-34\,b & 27.80 & 0.97 & 0.28 & 0.16--0.47 & 9 & \citet{welsh_transiting_2012}\\
  Kepler-35\,b & 20.73 & 0.91 & 0.40 & 0.18--0.49 & 7 & \citet{welsh_transiting_2012}\\
  Kepler-38\,b & 18.80 & 0.26 & 0.05 & 0.35--0.84 & 13 & \citet{orosz_neptune-sized_2012}\\
  Kepler-47\,b & 7.45 & 0.35 & 0.08 & 0.12--0.37 & 25 & \citet{orosz_kepler-47_2012,orosz_discovery_2019}\\
  Kepler-47\,c & 7.45 & 0.35 & 0.17 & 0.18--0.47 & 4 & \citet{orosz_kepler-47_2012,orosz_discovery_2019}\\
  Kepler-47\,d & 7.45 & 0.35 & 0.16 & 0.10--0.20 & 7 & \citet{orosz_kepler-47_2012,orosz_discovery_2019}\\
  Kepler-64\,b & 20.00 & 0.27 & 0.11 & 0.45--0.51 & 10 & \citet{kostov_gas_2013,schwamb_planet_2013}\\
  Kepler-413\,b & 10.12 & 0.66 & 0.17 & 0.12--0.39 & 9 & \citet{kostov_kepler-413b_2014}\\
  Kepler-453\,b & 27.32 & 0.20 & 0.50 & 0.29--0.57 & 3 & \citet{welsh_kepler_2015}\\
  Kepler-1647\,b & 11.26 & 0.80 & 0.22 & 0.2--0.37 & 3 & \citet{kostov_kepler-1647b_2016}\\
  Kepler-1661\,b & 28.16 & 0.31 & 0.21 & 0.31--0.49 & 3 & \citet{socia_kepler-1661_2020}\\
  TOI-1338\,b & 14.61 & 0.29 & 0.25 & 0.27--0.52 & 13$^\ddagger$ & \citet{kostov_toi-1338_2020}\\
  TIC 172900988\,b & 19.70 & 0.97 & 0.30 & 0.35--0.48 & 2 & \citet{kostov_tic_2021}\\
  \hline
  \multicolumn{7}{p{0.475\linewidth}}{\footnotesize
  $^\dagger$ The mean transit depth for transits across the primary star.\newline
  $^\ddagger$ Number of observed transits up to the end of Sector 98.
  }
 \end{tabular}
\end{table*}

Before confirming this event as a bona-fide CBP transit, there are several false positive scenarios that must be ruled out. Firstly, there is a possibility that the system is a single star and the $\approx2.2$ per cent eclipses are due to a planet. TIC v8.2 provides a radius of $2.86431\pm0.163317~R_{\odot}$ for this target. Therefore, from the eclipse depth, we obtain a radius estimate of $\approx4~R_{\text{J}}$. Such a large radius is unlikely to be planetary. This implies that either TIC 319011894 is a single star and the eclipses are due to a planet, but the radius given in the TIC is anomalously large, or TIC 319011894 is a stellar binary with a period of 8.3\,d, which would also be consistent with an inaccurate TIC radius~\citep{Stassun2019}. The \textit{Gaia} re-normalised unit weight error (RUWE) for this target is $1.03$, significantly below the commonly used threshold for unresolved binaries of 1.4, and it is not listed in the \textit{Gaia} Non-Single Star catalogue. Spectroscopic follow-up would quickly reveal if this system is a stellar binary or not.

\begin{figure}
 \includegraphics[]{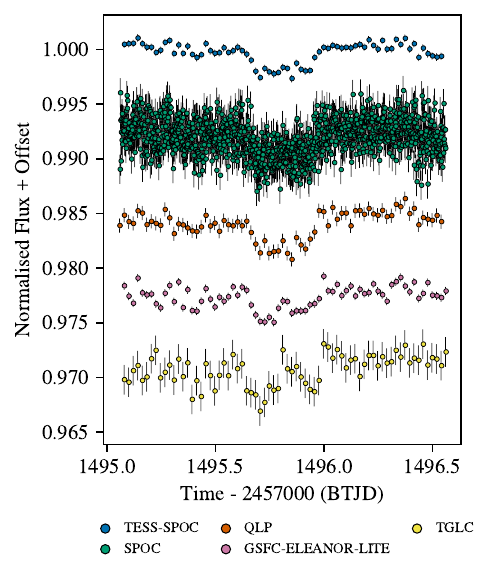}
 \caption{Candidate CBP transit in the Sector 7 light curve of TIC 319011894 extracted from five different \tess\ data reduction pipelines (top to bottom): TESS-SPOC~\citep[blue;][]{Caldwell2020}, SPOC~\citep[green;][]{jenkins_tess_2016}, QLP~\citep[orange;][]{Huang2020a,Huang2020b}, eleanor~\citep[pink;][]{Feinstein2019}, and TGLC~\citep[yellow;][]{Han2023}. For each light curve, the raw/SAP flux is shown. The candidate was identified from the TESS-SPOC light curve and is also clearly present in the SPOC, QLP, and eleanor light curves. The TGLC light curve (PSF flux) shows much more scatter than the other light curve products, but the presence of the candidate transit is consistent within the measurement uncertainties.}
 \label{fig: candidate light curve comparison}
\end{figure}

The 8.3\,d signal, or indeed the additional event, may originate from a nearby star or a background object, as several stars contaminate the SPOC aperture (each with $\Delta T\geq2.3$). However, there is no clear evidence for centroid shifts during the eclipses, and examining the light curves of each pixel in the TPF reveals that the signals are strongest on the pixels surrounding the target star. The 8.3\,d and candidate signals appear in the SPOC~\citep{jenkins_tess_2016}, Quick Look Pipeline~\citep[QLP;][]{Huang2020a,Huang2020b}, and eleanor~\citep{Feinstein2019} light curves available for this target on MAST. The candidate signal is less clearly present in the TESS-Gaia Light Curve~\citep[TGLC;][]{Han2023} PSF photometry, as its depth is similar to the scatter of the light curve, but the presence of the candidate transit is consistent given the measurement uncertainties. Figure~\ref{fig: candidate light curve comparison} illustrates this comparison between \tess\ data products for the candidate transit event. Ground-based photometric follow-up using detectors with a smaller pixel scale could help to confirm whether these events are on-target. 

Finally, there is a possibility that the candidate event is indeed a secondary eclipse of a stellar binary, but the other observable secondary eclipse (top-right panel of Fig.~\ref{fig: candidate event}) is affected by an instrumental systematic. The top-right panel of Fig.~\ref{fig: candidate event} shows where an additional event would occur if the candidate event was a secondary eclipse. To investigate this possibility, we detrended the photometry using different combinations of Cotrending Basis Vectors (CBVs) supplied by the SPOC. Firstly, we applied all MultiScale/SingleScale and Spike CBVs to the TESS-SPOC SAP light curve and used a grid of L2-norm regularisation terms from $10^{-4}-10^4$. Each light curve was then visually inspected. If there is a secondary eclipse at $\sim1512.5$ BTJD that is affected by an instrumental systematic, the systematic may be removed using a different set of CBVs than those that were used to produce the PDCSAP light curve. We find that there is no clear signal for any combination of CBV types and L2-norm regularisation terms.

\section{Application to the known transiting CBPs}
\label{sec: known CBPs}

We applied \monocbp\ to the known sample of transiting CBPs, including those discovered from the \kepler\ mission, allowing us to test the sensitivity of our approach to real CBP transit signatures. This provides a sample of 14 transiting CBPs with varying transit depth and duration, with different stellar hosts and photometric noise properties. Table~\ref{tab: known CBPs} shows some properties of the known transiting CBPs and their host stars.

\begin{figure*}
 \includegraphics[]{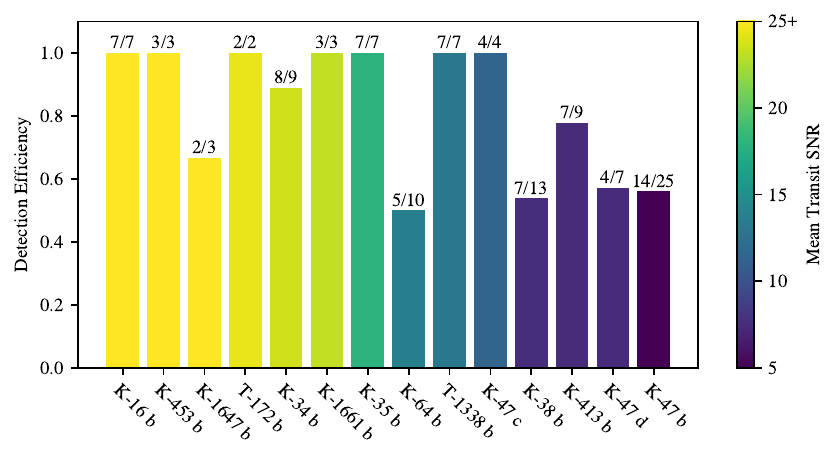}
 \caption{The detection efficiency (i.e., the proportion of individual transits identified by the framework) of \monocbp\ for the known transiting CBPs. The bars are colour-coded based on the mean SNR of an individual transit. For SNR~$\gtrsim10$ (the majority of the known planets), the framework identifies nearly all transits. Below SNR~$\lesssim8$, the framework struggles to identify all individual transits consistently, but in all cases sufficient transits are identified to confirm each planet. Note that Kepler-16\,b and Kepler-453\,b have much higher mean transit SNR than any of the other systems ($\approx470$ and $\approx90$, respectively).}
 \label{fig: detection efficiency known CBPs}
\end{figure*}

We note that for the 12 known transiting CBPs that were discovered by \kepler, we used the \kepler\ photometry to test {\tt mono-cbp}\footnote{The other two published transiting CBPs, TOI-1338\,b~\citep{kostov_toi-1338_2020} and TIC 172900988\,b~\citep{kostov_tic_2021}, were discovered by \tess.}. We did not use \tess\ data for the \kepler\ systems for two reasons. First, due to the faintness of the \kepler\ CBP host binaries ($T>13$ for all systems except Kepler-16) and the reduced photometric precision of \tess\ compared to \kepler, \tess\ light curves of these systems exhibit significantly higher noise levels, rendering any transits of these planets undetectable even if they occur. Second, many of these CBPs are no longer in a transiting configuration due to nodal precession of their orbits~\citep{Martin2017,socia_kepler-1661_2020}, meaning no transits would be observed regardless of photometric quality. Even if \tess\ were observing these systems during epochs when transits occurred, the unfavourable noise characteristics would still preclude detection.

For each \kepler\ system, we downloaded all available Quarters of long-cadence data from the MAST using {\tt lightkurve}, choosing to use the SAP photometry for the same reasons as outlined in Section~\ref{sec: data}. Each Quarter was split into 4 equal-length light curves such that the time baseline of each light curve was roughly the same length as a \tess\ sector. To perform the eclipse masking step for these systems, the ephemerides and eclipse parameters for each \kepler\ system were obtained from the \kepler\ Eclipsing Binary Catalogue~\citep{prsa_kepler_2011,kirk_kepler_2016}. Note that, since PH1/Kepler-64 is not present in this catalogue, we used the ephemeris from~\citet{schwamb_planet_2013} and determined the eclipse parameters by visual inspection.

Figure~\ref{fig: detection efficiency known CBPs} shows the detection efficiency (i.e., the number of individual transits that are identified by the detection framework) for the known transiting CBPs. There is a clear dependence on the mean SNR of the CBP transits for the detection efficiency of our framework; for mean transit SNR $\gtrsim10$, the detection framework identifies nearly all transits, while for SNR $\lesssim8$ (Kepler-38\,b, Kepler-413\,b, Kepler-47\,d, and Kepler-47\,b), our detection framework is less consistent, with detection efficiencies of $\lesssim80$ per cent. However, even for these lower SNR transits, we detect $\ge4$ transits in each case. We note that for some CBPs~\citep[e.g., Kepler-413\,b;][]{kostov_kepler-413b_2014}, the planet precesses in and out of transitability over time. In such cases, \monocbp\ has an advantage over detection methods that fold transits together, as it does not require transits at every conjunction and therefore does not lose SNR by stacking noise onto the folded transits during non-transiting epochs.

While the detection efficiency generally follows the expected SNR dependence, there are two exceptions to this trend. Firstly, only 2/3 of the Kepler-1647\,b transits were identified. One of the transits is a syzygy, and only the transit egress remains after eclipse masking. In the SAP photometry, there is a systematic increase in scatter after this transit, which causes the detection algorithm to identify the scatter as a TCE while the transit egress goes undetected. Secondly, PH1/Kepler-64\,b has a low detection efficiency, despite its many, moderately high SNR transits. This is likely due to the fact that the transit durations, which have a median of $\sim0.5$\,d, are on roughly the same timescale as the stellar variability present in the light curve. This combination can cause the detrending to remove the transit signal (see Section~\ref{subsec: injection-recovery}).

\section{Detection limits}
\label{sec: detection}

By basing our transit detection framework on identifying single-transit events, we implicitly bias our search to high SNR events, corresponding to larger CBPs. Through our initial tests on the known CBPs (Section~\ref{sec: known CBPs}), we have shown that \monocbp\ can identify many transits of the smallest of the known transiting CBPs in \kepler\ data (e.g., Kepler-47\,b), but this does not take into account the noise properties of the light curves in our sample, which are different to \kepler\ light curves. Here, we outline a series of tests to characterise the detection efficiency of \monocbp\ for a range of transit morphologies within our TEBC sample.

\subsection{Setup}
\label{subsec: setup}

In order to test the effectiveness of our search, we injected transit profiles across a grid of depths and duration into light curves from our sample (as defined in Section~\ref{sec: data}) and attempted to recover them using the same methodology as used for our search.

\begin{figure*}
 \includegraphics[]{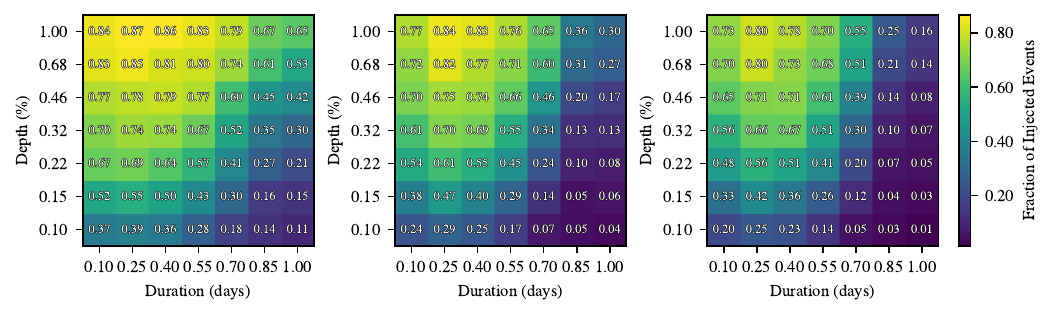}
 \caption{Results of our injection-recovery tests. For each square in the duration-depth grid, 1000 transit profiles with the corresponding depth and duration were injected into random light curves in our sample. The proportion of injections that were recovered by \monocbp\ are visualised by the colour bar and by the text in each grid square. \textbf{Left:} Recovery rates before we apply our vetting metrics (i.e., injected events that were identified by the transit search). \textbf{Middle:} Recovery rates after the default metrics are applied. \textbf{Right:} Recovery rates for injected events that are classified as T or AT by the model comparison. Generally, we achieve high recovery rates for deeper, shorter duration transits. Our vetting metrics and model comparison do have the unintended effect of filtering out some injected transit signals, particularly for the longer duration transits.}
 \label{fig: identified and recovered}
\end{figure*}

We generated transit profiles using {\tt exoplanet}~\citep{Foreman-Mackey2021} with a linear transit duration grid from 0.1--1\,d with a step size of 0.15\,d, and a base-10 logarithmic grid in transit depth from 0.1 to 1 per cent. The depth grid was chosen to be logarithmic such that shallower transits are sampled more finely, allowing us to more effectively quantify the lower limit of \monocbp\ with respect to transit depth.

Before injecting these transit signals, we first inverted the light curves such that we did not bias the results of our injection-recovery tests with signals that may be real. The detrending should be unaffected by this inversion since the timescales of trends in the light curves are the same, but there will be a change to the residual detrending artefacts that are identified, since the inversion process inverts the direction of systematic trends.

These transit profiles, with a total of 49 transit depth/duration combinations (i.e., a $7\,\times\,7$ grid), were then injected into the inverted light curves at a random time between the beginning and end of the \tess\ sector, ensuring that the transit is, in principle, detectable by requiring that the transit mid-time does not fall into a data gap (at least half the duration of the transit is within the data). For each transit depth/duration combination, 1000 transit profiles were injected, leading to a total of 49000 injection-recovery tests.

These light curves were then passed through \monocbp, applying the same vetting metrics as in Section~\ref{subsec: vetting}. A transit signal was considered ``identified'' if the time of the event was determined to within half the duration of the injected transit event, and ``recovered'' if it passed all of the automated vetting steps and the model comparison step (i.e., was classified as T/AT).

\subsection{False positive rate}
\label{subsec: FPR}

To estimate the false positive rate for our search, which we define as the number of false positive detections normalised to the number of light curves, we performed the steps in Section~\ref{subsec: setup} without the transit injection, allowing us to separate false positive detections from the (synthetic) true positives.

We note that our determination of the false positive rate is sensitive to the particular dataset used, since light curves with higher intrinsic noise will result in higher false positive rates. Therefore, the value that we report for the false positive rate is only valid in the particular case of our subsample of the TEBC. However, it is still informative for the purpose of characterising \monocbp\ for \tess\ SAP data with a variety of EB light curve properties, both astrophysical and instrumental in nature.

By applying this to our full sample of 3808 light curves and using the default threshold and detrending, we identified 3191 events. This translates to a false positive identification rate of $\approx0.84$ spurious events per light curve for our inverted sample. This is a rather high false positive rate, motivating the need for filtering using automated methods.

Applying our default vetting cuts and metrics (see Section~\ref{subsec: vetting}) removed $>70$ per cent of these false positives, resulting in 945 FPs being passed to the model comparison stage.

Finally, after we performed the model comparison, we obtain only 28 T candidates, with 520 AT candidates. Therefore, for the T candidates, we estimate our FP rate to be $\approx0.007$ false positives per light curve. Considering both T and AT candidates, this value increases to $\approx0.14$ false positives per light curve. While this transit search method identifies a large proportion FPs, our automated vetting is effective at reducing the amount of false positive detections.

\subsection{Results of transit injection-recovery}
\label{subsec: injection-recovery}

We carried out the steps laid out in Section~\ref{subsec: setup}. Our ``identified'' recovery rate for each transit morphology is shown in the left panel of Fig.~\ref{fig: identified and recovered}.

The identification rates show that, as expected, \monocbp\ is better at identifying deeper transits ($\geq0.46$ per cent) with intermediate duration ($0.25-0.55$\,d). \monocbp\ is least sensitive to shallow, long duration transits; our detrending causes these longer-duration transits to be removed before the transit search. On the short-duration end, events with durations of 0.1\,d are more difficult to identify than events that are $0.25-0.4$\,d in duration; at 30 minute cadence, this corresponds to only 4 or 5 cadences. Therefore, whether an event passes the detection criterium of 3 consecutive cadences beneath our threshold is more dependent on the noise properties of the light curve. Lastly, for a given transit depth, the identification rate decreases for longer durations. This highlights how much the light curve detrending affects the detectability of CBP transits with long durations.

\begin{figure}
 \includegraphics[]{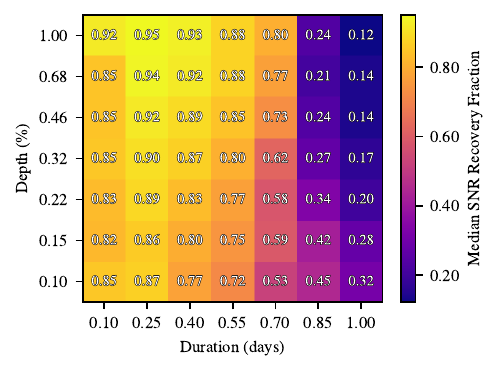}
 \caption{Recovered Signal-to-Noise Ratio (SNR) of injected transit signals as a function of transit depth and duration, before our vetting metrics are applied. Most of the signal is retained for transits with durations of $\leq0.7$\,d, while our detrending approach removes a significant proportion of the signal for longer duration transits.}
 \label{fig: SNR recovery}
\end{figure}

To further investigate the effects of our detrending, we calculated the SNR of each injected transit using Eq.~\ref{eq: SNR} and the injected depth and duration and compared these to the SNR using the recovered depth and duration. The results are shown in Fig.~\ref{fig: SNR recovery}. As expected, the majority of the injected transit SNR is recovered by \monocbp\ for transits with the highest recovery rates. There is also a steep decrease in the recovered SNR for transits with durations $\geq0.85$\,d, consistent with our lower identification rates at these longer durations.

\begin{figure}
 \includegraphics[]{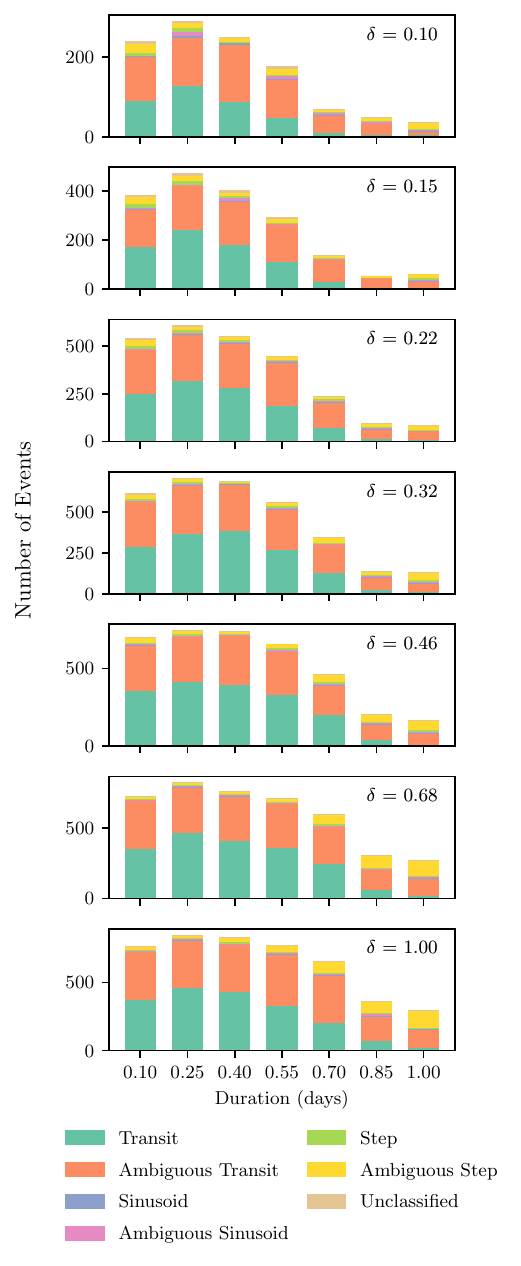}
 \caption{Model comparison classifications of the recovered transits from the injection-recovery tests, after applying the default vetting cuts. Most are classified as transits/ambiguous transits. Many of the longer duration transits are classified as steps due to the detrending altering the transit morphology.}
 \label{fig: injection-retrieval model comparison}
\end{figure}

It is certainly possible that the automated vetting metrics that we apply to our candidate events remove real transit signals. To assess the impact of this part of the framework, we applied the standard cuts and metrics to our ``identified'' sample to see how this affected the recovery rate. Note that since we used a random subset of our light curve sample (1000 light curves out of 3808), the number of correlated events used to calculate our Skye metric flag will be different to the run without injected events. To be conservative in our recovery rate estimates, we used the Skye times (i.e., the times corresponding to events that are flagged by the Skye metric) from the false positive rate estimation (see Section~\ref{subsec: FPR}), which used all 3808 light curves.

The recovery rate of the injected transits after we apply our vetting metrics are shown in the middle panel of Fig.~\ref{fig: identified and recovered}. Our vetting metrics remove a small proportion of injected transits for deeper, intermediate-duration transits (typically $\lesssim10-15$ per cent of the total injected events) but significantly reduce the recovery rate for transits with long durations, with as much as 67 per cent of transits being removed for the $\delta=0.1$~per cent, $t_\text{dur}=1$\,d case. This is an important caveat to consider when searching for such long duration, shallow events with \monocbp; these transits are affected mostly by the detrending-dependence cut, since they are removed by the shortest window lengths in our biweight grid (Section~\ref{subsubsec: biweight}). The effect of each individual cut is discussed in Appendix~\ref{app: individual vetting cuts}.

Lastly, we applied our model comparison procedure to the events which passed our automated vetting steps. For each event, we use the default vetting cuts and metrics and pass only the events which were injected (i.e., only events which are coincident with the time of the injected event) to the model comparison. We then classify them according to our AIC and RMSE thresholds as set out in Section~\ref{subsec: vetting}. The recovery rates of events that were classified as T or AT are shown in the right panel of Fig.~\ref{fig: identified and recovered}. There is a small decrease in the recovery rate for each depth/duration combination (typically of order a few per cent), apart from the deeper, longer-duration events, whose morphologies are the most effected by the detrending process.

Figure~\ref{fig: injection-retrieval model comparison} shows the proportion of each model comparison classification as a function of injected transit depth and duration. Most events are classified as T/AT, particularly for transits with $t_\text{dur}\leq0.7$\,d. This shows that the model comparison works effectively at correctly classifying the morphology of the injected transits. It is also apparent that, for longer duration transits of a given depth, a larger proportion of the injected transits are misclassified, with many of these events being best fit by the step model. As mentioned previously, this is due to the detrending process distorting the shape of these transit profiles.

\begin{table*}
  \centering
  \caption{Detection efficiency as a function of transit depth for two different binary host configurations. Planet radii are calculated from Eq.~\ref{eq: dilution}. The values reported for the detection efficiency and their uncertainties were calculated by taking the mean and standard deviation of each row on the right-hand panel of Fig.~\ref{fig: identified and recovered}}
  \label{tab: detection efficiency}
  \begin{tabular}{c|cc|cc}
    \hline
    $\delta$ (\%) & \multicolumn{2}{c|}{Planet Radius ($R_{\rm J}$)} & \multicolumn{2}{c}{Detection Efficiency} \\
     & G+M ($q<<1$) & G+G ($q=1$) & Duration-averaged & $t_{\rm dur} < 0.85$\,d \\
    \hline
    0.10 & 0.31 & 0.44 & $0.13\pm0.09$ & $0.17\pm0.07$ \\
    0.15 & 0.38 & 0.53 & $0.22\pm0.15$ & $0.30\pm0.10$ \\
    0.22 & 0.46 & 0.65 & $0.33\pm0.20$ & $0.43\pm0.13$ \\
    0.32 & 0.55 & 0.78 & $0.41\pm0.23$ & $0.54\pm0.13$ \\
    0.46 & 0.67 & 0.93 & $0.47\pm0.25$ & $0.61\pm0.12$ \\
    0.68 & 0.80 & 1.13 & $0.54\pm0.24$ & $0.68\pm0.10$ \\
    1.00 & 0.97 & 1.38 & $0.57\pm0.24$ & $0.71\pm0.09$ \\
    \hline
  \end{tabular}
\end{table*}

\section{Discussion}
\label{sec: discussion}

\subsection{Sensitivity of \monocbp\ to planet radius}
\label{subsec: constraints}

In Section~\ref{sec: detection}, we characterised the performance of \monocbp\ when applied to a sample of stars from the TEBC (defined in Section~\ref{sec: data}). Using the results from Section~\ref{sec: detection}, we can place some constraints on the size of transiting CBPs that may be present in the data.

Constraining exoplanet populations from monotransits is challenging, particularly for CBPs. For planets orbiting single stars, the transit depth, duration, and stellar radius can be used to estimate the planet's orbital period and radius. For CBPs, however, transit duration depends not only on the planet's orbital period but also on the binary's orbital phase at conjunction and three-body dynamical interactions. Given the information from \monocbp\ and our injection-recovery tests, it is therefore difficult to directly relate observables from monotransits to physical planet properties for CBPs, as many factors, particularly the properties of the stellar binaries, remain unknown for most systems. We can, however, estimate mono-cbp's sensitivity to various planet radii for our light curve sample using representative binary configurations.

Since the sample of binary stars in our search are much too close in orbital separation to be resolved by \tess, the light from the secondary star will dilute the transit depth of a CBP transiting the primary star, and vice versa. Equation 11 from~\citet{Martin2019} states that the transit depth of a transiting CBP depends on the mass ratio of the host binary, $q_{\rm bin}$, as follows:

\begin{equation}
    R_{\text{CBP}}=R_A\sqrt{\delta(1+q_{\rm bin}^{3.5})},
    \label{eq: dilution}
\end{equation}

\noindent where $R_{\text{CBP}}$ is the radius of the CBP, $R_A$ is the radius of the primary star (or the star being transited), and $\delta$ is the observed transit depth.

Consider a host binary with the median brightness of our sample ($T\approx10.2$, see the top panel of Fig.~\ref{fig: EB sample}) that consists of a solar twin ($M_A=1M_{\odot}; R_A=1R_{\odot}$) primary star and a low-mass secondary star, such that $1+q_{\rm bin}^{3.5}\approx1$ and there is negligible dilution from the secondary star for transits across the primary star. From the right panel of Fig.~\ref{fig: identified and recovered}, we can estimate the detection efficiency for planets of a given depth (radius) across all sampled transit durations, as well as transit durations which correspond to the population of the known transiting CBPs (fifth column of Table~\ref{tab: known CBPs}). These results are shown in Table~\ref{tab: detection efficiency}.

For example, \monocbp\ obtains a duration-averaged (i.e., $0.1\,{\rm d}\leq t_{\rm dur}\leq1.0$\,d) detection efficiency of $0.33\pm0.20$ for CBP transits across the primary star for $R_{\rm CBP}\approx0.46~R_{\text{J}}$ (corresponding to a transit depth of 0.22 per cent). Similarly, we estimate a duration-averaged detection efficiency of $0.47\pm0.25$ for $R_{\rm CBP}\approx0.66~R_{\text{J}}$ (corresponding to a transit depth of 0.46 per cent). If we exclude the region of parameter space where a significant reduction in the recovered SNR occurs (i.e., $t_{\rm dur} \geq 0.85$\,d), thereby restricting to a duration range that encompasses the distribution of transit durations for the known transiting CBPs (Table~\ref{tab: known CBPs}), \monocbp\ achieves detection efficiencies of $0.61\pm0.12$ and $0.43\pm0.13$ for $R_{\rm CBP}\approx0.67~R_{\text{J}}$ and $R_{\rm CBP}\approx0.46~R_{\text{J}}$, respectively. The improved detection efficiencies in the restricted duration range demonstrate that \monocbp\ can reliably detect sub-Saturn-sized CBPs (a detection efficiency of $>50$ per cent) in \tess\ data with configurations similar to the known systems, where transit durations remain below the threshold at which the detrending significantly degrades the recovered SNR (Fig.~\ref{fig: SNR recovery}).

Now consider the scenario where a CBP transits the primary star of an equal-mass stellar binary with the median brightness of our sample, each star being a solar twin ($M_{A,B}=1M_{\odot}$; $R_{A,B}=1R_{\odot}$; $1+q_{\rm bin}^{3.5}=2$). This scenario corresponds to the maximum dilution of the depth of a CBP transit, and hence the maximal decrease in detection sensitivity in terms of planet radius, as shown in the third column of Table~\ref{tab: detection efficiency}. For example, in this configuration, \monocbp\ is able to consistently ($>50$ per cent of the time) detect transits of Jupiter-sized planets for $0.1\,{\rm d} \leq t_{\rm dur} \leq 1.0\,{\rm d}$, while for $t_{\rm dur} \leq 0.85$\,d, \monocbp\ consistently detects Saturn-sized CBP transits. Since equal-mass binaries represent the most challenging configuration in terms of dilution and range of geometrical TDVs, these detection efficiencies establish a conservative lower bound on \monocbp's detection sensitivity across the full range of binary mass ratios.

An important distinction should be noted between the detection of individual transits and the detection of a transiting CBP. Typically, the detection of a CBP requires multiple transits, as well as some physically motivated constraints on the system which confirm it as a circumbinary body (e.g., TTVs/TDVs, quasi-periodicity, and dynamical restrictions). \kepler, with its $4$-year baseline of continuous photometry, was more suited to identifying multiple transits of these necessarily long-period planets. Therefore, given \tess's more intermittent photometry and the limited characterisation of most EBs in our sample, individual transit detections serve primarily to identify promising CBP candidates for follow-up, such as TIC 319011894, rather than confirmed systems. Nevertheless, the detection sensitivities established here provide a validated framework for systematic CBP searches in \tess\ data, which we discuss in Section~\ref{subsec: future}.

\subsection{Future applications and improvements}
\label{subsec: future}

The validation of \monocbp\ on the TEBC sample opens several avenues for future work, including: application to larger \tess\ datasets, follow-up strategies for CBP candidates, methodological improvements to increase detection efficiency, and adaptation to future transit surveys.

\textbf{Expanding the \tess\ search}. We intend to apply \monocbp\ to additional \tess\ datasets not included in this work. Beyond the TESS-SPOC light curves analysed here, alternative data reduction pipelines (e.g., QLP) provide photometry for our TEBC sample that remains to be searched. Furthermore, hundreds of thousands of EBs have been identified in the \tess\ FFIs, many with calculated ephemerides~\citep[e.g.,][]{Kostov2025}, providing a substantial dataset for future systematic CBP searches. The \tess\ General Investigator programme has also provided observations for EBs at 2-minute cadence, adding to the sample of potential CBP hosts to analyse with \monocbp.

\textbf{Follow-up of TIC 319011894}. We are exploring several options to confirm or rule out the candidate transit event as originating from a CBP. Ground-based photometric follow-up during predicted conjunction windows of the primary eclipses will reduce the uncertainty on the binary ephemeris, and additional photometric monitoring covering the orbital phase where a secondary eclipse would occur will help to rule out the possibility that the candidate event is itself a secondary eclipse. Also, we plan to obtain spectroscopic follow-up in order to confirm the binary nature of the target and to further constrain the configuration of the system.

\textbf{Methodological improvements}. From Section~\ref{sec: detection}, it is clear that there is room for improvement when it comes to removing false positives while retaining real transit signatures. With this goal in mind, we plan to explore alternative detrending methods to increase our detection efficiency for longer-duration transit events, as well as making improvements to the automated vetting procedure. Additionally, the main bottleneck in terms of computational efficiency is the model comparison vetting, and so we aim to further optimize this component of the framework.

\textbf{Application to future missions}. Beyond \tess, \monocbp's framework is well-suited to future space-based transit surveys. \textit{PLATO} will provide high-quality photometry with an observational baseline of at least 2 years for a 2132 deg$^{2}$ region of the sky~\citep{Rauer2025}. This observational strategy is more amenable to the detection of long-period transiting exoplanets, a category encompassing the current sample of transiting CBPs. \monocbp\ could be adapted to a sample of EBs observed by \textit{PLATO}, aiding in the identification of new transiting CBPs as the data are released, as well as identifying CBPs which may precess into transiting configurations during the mission lifetime.

\section{Conclusions}
\label{sec: conclusions}

In this paper, we have developed and validated \monocbp, a framework for identifying single-transit events in eclipsing binary light curves, and applied it to search for circumbinary planets in a subsample of the \tess\ Eclipsing Binary Catalogue~\citep{prvsa_tess_2022}. For our sample of 512 EBs, we masked the stellar eclipses, applied a custom detrending procedure, searched for threshold-crossing events, and performed a series of automated vetting checks before visually inspecting events that satisfied our vetting criteria.

We identified one candidate transit event in the Sector 7 light curve of TIC 319011894 (Fig.~\ref{fig: candidate event}). Given the current data, this event appears to be on-target and astrophysical in origin, though further observations are required to confirm its true nature and investigate the system's properties.

To characterize \monocbp's performance, we applied the framework to light curves of known transiting CBP hosts, successfully identifying $\geq50$ per cent of the detectable transits for each planet, reaching $>75$ per cent for 9 of the 14 planets (Fig.~\ref{fig: detection efficiency known CBPs}). We then performed comprehensive injection-recovery tests using simulated transit events with varying depths and durations injected into the \tess\ light curves of our EB sample. These tests revealed that the choice of vetting metrics significantly affects the recovery of injected transits. Therefore, caution must be exercised when selecting vetting thresholds for automated transit searches with \monocbp.

Our injection-recovery tests established \monocbp's detection sensitivity across a range of planet sizes and binary configurations. For binary systems with the median brightness of our sample ( $T\approx10.2$) and low secondary mass ratios ($q_{\rm bin} << 1$), \monocbp\ achieves detection efficiencies of $0.61\pm0.12$ for sub-Saturn-sized planets ($R_{\rm CBP}\approx0.67R_{\rm J}$) and $0.43\pm0.13$ for planets slightly larger than Neptune ($R_{\rm CBP}\approx0.46R_{\rm J}$) when restricting to transit durations typical of known transiting CBPs ($t_{\rm dur}\leq0.85$\,d). For equal-mass binary configurations ($q_{\rm bin}\sim1$), \monocbp\ maintains sensitivity to Jupiter-sized planets across the full tested duration range and Saturn-sized planets for shorter-duration transits. These detection limits establish a conservative baseline for \monocbp's capabilities across a range of binary mass ratios.

While individual transit detections cannot definitively confirm CBPs without additional information, \monocbp\ provides a validated framework for systematically identifying promising candidates in large samples of EBs. The methodology demonstrated here can be extended to the hundreds of thousands of EBs identified in \tess\ FFIs and adapted to future missions such as PLATO, which will be better suited to detecting multiple transits of CBPs. As the sample of confirmed transiting CBPs grows, more concrete constraints will be able to be made on the population of giant CBPs, addressing fundamental questions about planet formation in binary environments.

\section*{Acknowledgements}

We would like to thank the referee for their detailed feedback, which greatly improved the quality of this paper. Additionally, the authors have several people to thank for helpful discussions which aided in forming the direction of this research, including David Armstrong, Faith Hawthorn, David Martin, Dominic Oddo, and Amaury Triaud.

This paper includes data collected by the \tess\ mission. Funding for the \tess\ mission is provided by the NASA's Science Mission Directorate. This paper also includes data collected by the \kepler\ mission and obtained from the MAST data archive at the Space Telescope Science Institute (STScI). Funding for the \kepler\ mission is provided by the NASA Science Mission Directorate. STScI is operated by the Association of Universities for Research in Astronomy, Inc., under NASA contract NAS 5–26555.

This research made use of Lightkurve, a Python package for \kepler\ and \tess\ data analysis~\citep{lightkurve_2018}. This work also made use of Astropy:\footnote{http://www.astropy.org} a community-developed core Python package and an ecosystem of tools and resources for astronomy \citep{astropy:2013, astropy:2018, astropy:2022}. This research made use of {\tt exoplanet} \citep{Foreman-Mackey2021,exoplanet:zenodo} and its dependencies \citep{exoplanet:agol20,
exoplanet:arviz, exoplanet:luger18}.

This research has made use of the NASA Exoplanet Archive, which is operated by the California Institute of Technology, under contract with the National Aeronautics and Space Administration under the Exoplanet Exploration Program.

The authors would like to acknowledge the University of Warwick Research Technology Platform (Scientific Computing) for assistance in the research described in this paper.

BDRD is supported by a STFC studentship (Ref: 2881369). DJAB is supported by the UK Space Agency. SG has been supported by STFC through consolidated grants ST/P000495/1, ST/T000406/1 and ST/X001121/1. JRF is supported by the ERC/UKRI Frontier Research Guarantee programme (EP/Z000327/1/CandY).

%%%%%%%%%%%%%%%%%%%%%%%%%%%%%%%%%%%%%%%%%%%%%%%%%%
\section*{Data Availability}

\monocbp\footnote{https://github.com/bdrdavies/mono-cbp} is publicly available on GitHub.

A live version of the TESS Eclipsing Binary Catalogue~\citep{prvsa_tess_2022}, the source of our sample, can be found at https://tessebs.villanova.edu/. 

%%%%%%%%%%%%%%%%%%%% REFERENCES %%%%%%%%%%%%%%%%%%

% The best way to enter references is to use BibTeX:

\bibliographystyle{mnras}
\bibliography{references}

% Alternatively you could enter them by hand, like this:
% This method is tedious and prone to error if you have lots of references
%\begin{thebibliography}{99}
%\bibitem[\protect\citeauthoryear{Author}{2012}]{Author2012}
%Author A.~N., 2013, Journal of Improbable Astronomy, 1, 1
%\bibitem[\protect\citeauthoryear{Others}{2013}]{Others2013}
%Others S., 2012, Journal of Interesting Stuff, 17, 198
%\end{thebibliography}

%%%%%%%%%%%%%%%%%%%%%%%%%%%%%%%%%%%%%%%%%%%%%%%%%%

%%%%%%%%%%%%%%%%% APPENDICES %%%%%%%%%%%%%%%%%%%%%

\appendix

\section{Effects of Vetting Cuts on Injection-Recovery Tests}
\label{app: individual vetting cuts}

As noted in Section~\ref{subsec: injection-recovery}, the automated vetting metrics implemented in \monocbp, while being effective at removing false positives, can remove realistic transit signatures. In this Appendix, we provide an outline of the effects of each individual vetting metric on the recovery rate of our injected transit profiles.

\begin{figure}
 \includegraphics[]{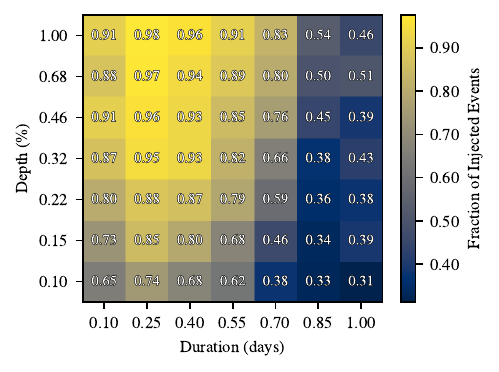}
 \caption{The fraction of identified transit profiles from the injection-recovery tests (Section~\ref{sec: detection}) that survive our default vetting metrics (Section~\ref{subsec: vetting}).}
 \label{fig: identified after cuts}
\end{figure}

Figure~\ref{fig: identified after cuts} shows the fraction of retained transit signatures after the default vetting metrics (excluding the model comparison) have been applied (i.e., the ratio of the left panel and the middle panel of Fig.~\ref{fig: identified and recovered}). This shows that $\gtrsim80$ per cent of identified transits survive the automated vetting for $\delta\geq0.22$ per cent, $t_{\text{dur}}\leq0.4$\,d. The vetting metrics also have the unintended effect of removing the majority of long-duration ($t_{\text{dur}}\geq0.85$\,d) transit signals.

\begin{figure}
 \includegraphics[]{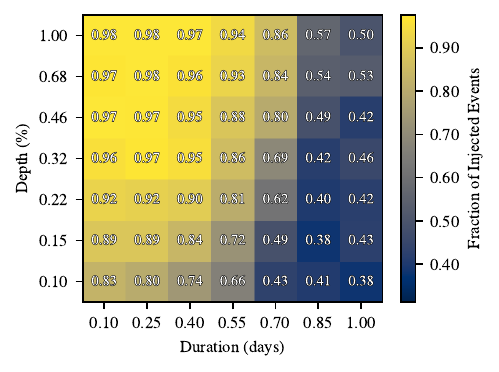}
 \caption{Same as Fig.~\ref{fig: identified after cuts}, but excluding the SNR cut.}
 \label{fig: identified after cuts no SNR}
\end{figure}

\begin{figure}
 \includegraphics[]{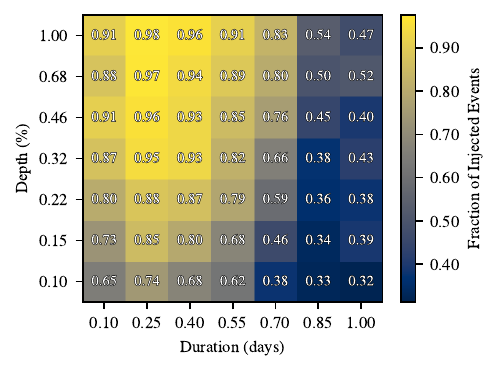}
 \caption{Same as Fig.~\ref{fig: identified after cuts}, but excluding the duration cut.}
 \label{fig: identified after cuts no duration}
\end{figure}

\begin{figure}
 \includegraphics[]{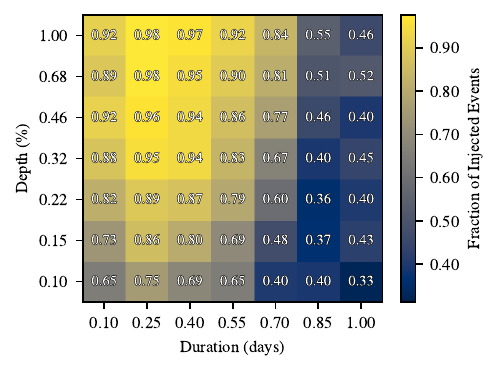}
 \caption{Same as Fig.~\ref{fig: identified after cuts}, but excluding the Skye metric flag cut.}
 \label{fig: identified after cuts no Skye}
\end{figure}

\begin{figure}
 \includegraphics[]{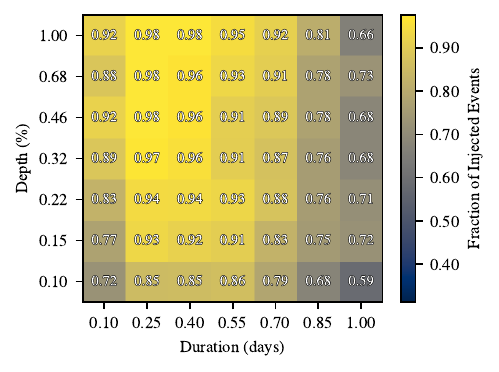}
 \caption{Same as Fig.~\ref{fig: identified after cuts}, but excluding the detrending-dependence flag cut.}
 \label{fig: identified after cuts no det dependence}
\end{figure}

Figures~\ref{fig: identified after cuts no SNR}--\ref{fig: identified after cuts no det dependence} show the effects of removing an individual cut from each stage of the automated vetting process (SNR cut, duration cut, Skye metric flag cut, and detrending-dependence flag cut, respectively). 

The SNR cut (Fig.~\ref{fig: identified after cuts no SNR}) has the largest effect on the short duration events, since such events have the smallest SNR for a given transit duration (SNR $\propto\sqrt{t_{\text{dur}}}$). This cut removes $\approx10-20$ per cent of injected transits with $t_{\text{dur}}=0.1$\,d, with similar false negative rates for $\delta=0.1$ per cent events.

The duration cut has a minor effect on the recovery rate, reducing the recovery rate for injected events with $t_{\text{dur}}=1$\,d by at most $\approx1$ per cent (Fig.~\ref{fig: identified after cuts no duration}). This occurs because \monocbp\ can occasionally overestimate the transit duration. As expected, this cut has no effect on injected events with $t_{\text{dur}}<1$\,d.

The Skye metric flag cut seems to be roughly independent of the injected transit depth and duration (Fig.~\ref{fig: identified after cuts no Skye}). The reduction in the number of recovered events is of order a few per cent across the grid. This is expected as this metric is only correlated with the time at which the transit occurs, which was randomly selected for our injection-recovery tests.

The detrending-dependence flag cut has the largest effect on the injected transit events, particularly those with longer durations (Fig.~\ref{fig: identified after cuts no det dependence}). An event is flagged as detrending-dependent if it is detected in $<18/21$ biweight-detrended light curves. The long-duration, shallow events are the most likely to be flagged since the biweight grid is applied with window lengths as low as 1\,d, meaning that such transits are susceptible to being fit by the detrending. Shallow events are also more likely to be flagged as detrending-dependent, since small changes in the detrending may cause the event to straddle the $3\,\times\,$MAD detection threshold, thus possibly evading detection in a proportion of the biweight-detrended light curves.

%%%%%%%%%%%%%%%%%%%%%%%%%%%%%%%%%%%%%%%%%%%%%%%%%%

% Don't change these lines
\bsp	% typesetting comment
\label{lastpage}
\end{document}